\documentclass[structabstract]{aa}  
\usepackage{graphicx}
\usepackage{txfonts}
\begin{document}
\title{Abundance analysis for long period variables}
\subtitle{Velocity effects studied with O-rich dynamic model atmospheres}

\author{T. Lebzelter
          \inst{1}
          \and
          W. Nowotny
          \inst{1}
          \and
          S. H\"ofner
          \inst{2}
          \and
          M.T. Lederer
          \inst{1}
          \and 
          K.H. Hinkle
          \inst{3}
          \and
          B. Aringer
          \inst{1,4}
          }

\institute{University of Vienna, Department of Astronomy, 
              T\"urkenschanzstra{\ss}e 17, A-1180 Wien, Austria\\
              \email{thomas.lebzelter@univie.ac.at}
          \and
             Department of Physics and Astronomy, 
             Division of Astronomy and Space Physics,
             Uppsala University,
             Box 515, SE-75120 Uppsala, Sweden
          \and
             National Optical Astronomy Observatories
             \thanks{Operated by the Association of Universities for
             Research in Astronomy, Inc. under cooperative agreement 
             with the National Science Foundation},
             950 N. Cherry Avenue, Tucson, Arizona 85726, USA
          \and
             Osservatorio Astronomico di Padova -- INAF,
             vicolo dell'Osservatorio 5, I-35122 Padova, Italy 
             }

\date{Received ; accepted }

\titlerunning{Abundance analysis for long period variables}
\authorrunning{T. Lebzelter et al.}

 
\abstract
{Measuring the surface abundances of AGB stars is an important tool for studying the effects of nucleosynthesis and mixing in the interior of low- to intermediate mass stars during their final evolutionary phases. The atmospheres of AGB stars can be strongly affected by stellar pulsation and the development of a stellar wind, though, and the abundance determination of these objects should therefore be based on dynamic model atmospheres.}
{We investigate the effects of stellar pulsation and mass loss on the appearance of selected spectral features (line profiles, line intensities) and on the derived elemental abundances by performing 
a systematic comparison of hydrostatic and dynamic model atmospheres.}
{High-resolution synthetic spectra in the near infrared range were calculated based on two dynamic 
model atmospheres (at various phases during the pulsation cycle) as well as a grid of hydrostatic COMARCS models with effective temperatures $T_{\rm eff}$ and surface gravities log\,$g$ over an adequate range. Equivalent widths of a selection of atomic and molecular lines (Fe, OH, CO) were derived in both cases and compared with each other.}
{In the case of the dynamic models, the equivalent widths of all investigated features vary over the pulsation cycle. A consistent reproduction of the derived variations with a set of hydrostatic models is not possible, but several individual phases and spectral features can be reproduced well with the help of specific hydrostatic atmospheric models. In addition, we show that the variations in equivalent width that we found on the basis of the 
adopted state-of-the-art dynamic model atmospheres agree qualitatively with observational results for the Mira R\,Cas over its light cycle.}
{The findings of our modelling form a starting point to deal with the problem of abundance determination in strongly dynamic AGB stars (i.e., long-period variables). Our results illustrate that some quantities such as the C/O ratio
can probably still be determined to a reasonable accuracy, but the measurement of other quantities will be 
hampered by the dynamics. The qualitative agreement with observations of R\,Cas opens promising possibilities for a forthcoming quantitative comparison of our synthetic spectra with observed ones of AGB variables in the globular cluster 47\,Tuc.}


\maketitle

\section{Introduction}

Measurements of abundances in atmospheres of highly evolved red giants contribute to the understanding of two major 
astrophysical aspects, namely the processes inside red giant stars such as nucleosynthesis and mixing, and, secondly, 
the enrichment of the interstellar medium by freshly produced elements due to the mass loss of these stars. The determination of accurate element abundances is complicated by the large number of spectral lines from neutral atoms and molecules that give rise to heavy blending, especially in the optical range of the spectrum. The problem is less severe in the near infrared range, which has become increasingly important for this kind of analysis. 

When a low or intermediate mass star evolves to the Asymptotic Giant Branch (AGB) phase, a second major problem occurs concerning the spectral analysis. Stars in this part of the HRD exhibit large amplitude radial pulsations. Being members of the variability class of so-called long period variables (LPVs), these objects change their visual brightnesses by up to several magnitudes over a few hundred days. The pulsating stellar interiors lead to complex 
velocity fields in the stellar atmospheres (e.g., Nowotny et al. \cite{NowAH09}) and subsequently to phase-dependent line profile variations (asymmetric and split lines or emission features) as seen in observed spectra of these stars (e.g.\,Hinkle et al.\,\cite{HHR82}). Thus, a hydrostatic description of these objects is inappropriate. However, it is this evolutionary phase where major nucleosynthesis and mixing processes take place, and therefore great interest in measuring stellar abundances exists.

In the past, a few attempts have been made to investigate the complicated profiles of individual lines observed in 
spectra of variable AGB stars on the basis of model calculations. Using the dynamic models for pulsating red giants 
without mass loss of Bessell \& Scholz (\cite{BessS89}), Scholz (\cite{Schol92}) showed how velocity fields in Mira 
photospheres distort line profiles of different species and affect measurements of equivalent widths and curves of 
growth for abundance analyses from these spectra. This fundamental paper, also referencing earlier works in this field, investigated the possibility to "predict accurately the equivalent width of lines which are little affected by outflow/infall velocities by means of a conventional hydrostatic model in place of a dynamical model" (Scholz \cite{Schol92}). He found that lines formed in deep layers below the arising shock fronts may be accessible to a classical curve of growth study, but that in some cases this approach seriously fails, so that abundance determination for large amplitude red variables remains a difficult problem.

In this context, a major challenge for dynamic model atmospheres was to reproduce the line profile variations observable in spectra of Mira type variables. First results concerning this were presented by Bessell et al. (\cite{BesSW96}), who calculated synthetic line profiles (CO, Fe\,I) based on the models of Bessell \& Scholz (\cite{BessS89}). They could qualitatively reproduce some of the observed line profile characteristics (asymmetries, shifted, or doubled lines).

McSaveney et al. (\cite{MWSLH07}) attempted to derive abundances from high resolution spectra of pulsating AGB stars in the Magellanic Clouds by using the same type of model atmospheres. They succeeded in fitting the spectra of some stars at selected phases during the pulsation cycle. However, they had to restrict themselves to cases of relatively smooth atmospheric velocity fields. They were unable to derive abundances for model phases with complex atmospheric structures (shock waves), which cause distortions in the line profiles (emission components, line splitting).

While the Australia-Heidelberg model atmospheres (e.g., Bessell \& Scholz \cite{BessS89}, Bessell et al. \cite{BesSW96}, Hofmann et al. \cite{HofSW98}) represent pulsating red giants without mass loss, the Berlin models (e.g. Fleischer et al. \cite{FleGS92}, Winters et al. \cite{WiFLS97}+\cite{Winters00}) describe the optically thick dusty outflows of evolved C-rich AGB stars with high mass-loss rates. Based on the latter, Winters et al. (\cite{Winters00}) presented synthetic fundamental and first overtone CO line profiles. By comparing reasonably with observed ones for the obscured \mbox{C-rich} Mira IRC+10216, these spectral lines can be used to investigate the radial structure and the temporal variation in the dusty envelope of these objects.

The non-grey dynamic model atmospheres of H\"ofner et al. (\cite{HoGAJ03}) represent a combination of both approaches 
and describe the outer layers of an AGB star from the pulsating deep photosphere (piston driven), to the dust forming 
region where the stellar wind is triggered and beyond to the region of the steady outflow (Nowotny et al. 
\cite{Nowo05a}). Keeping in mind the work of Scholz (\cite{Schol92}), we conducted a similar exercise to investigate the problem of deriving abundances for LPVs by utilising the advanced dynamic models of H\"ofner et al. (\cite{HoGAJ03}). 

The major aim of the work presented here was to study the effect of atmospheric dynamics (pulsation, stellar wind) on the intensity of spectral lines for typical AGB stars. For this purpose, we compared selected spectral features in synthetic spectra computed on the basis of the aforementioned dynamic atmospheric models (Sect.\,\ref{s:dynvsstat}) with the corresponding results based on hydrostatic model atmospheres (Sect.\,\ref{s:hydro}). In a second step, we investigate observed high-resolution spectra of a typical Mira to see if the effects we found by the modelling are 
also recognisable in the observational data (Sect.\,\ref{s:obsvsmod}).

\section{Methods}

The starting point for the calculations are atmospheric structures, either from hydrostatic or dynamic models. Based on this, we calculate synthetic spectra and derive equivalent widths for a selection of atomic and molecular features. In our study, we follow the approach of an observational abundance study: to have a starting point for fitting an observed spectrum with a synthetic one, basic quantities such as effective temperature are derived from observed colours and subsequently the observed target is compared with the corresponding model. In our case, we 
also use a typical colour, ($J$--$K$), as an indicator of $T_{\rm eff}$. That is, we investigate how line intensities deviate in the dynamical case from the hydrostatic model, both related by a given colour. In doing so, we assume that ($J$--$K$) correlates with $T_{\rm eff}$, keeping in mind that the term effective temperature may become questionable for the very evolved AGB stars featuring strong pulsations and mass loss (Baschek et al. \cite{BasSW91}, Nowotny et al. \cite{Nowo05b}).

\subsection{Model atmospheres}

\begin{table}
\begin{center}
\caption{Characteristics of the dynamic atmospheric models used for the modelling (P = pulsating, PM = pulsating and mass-losing). The notation was adopted from previous papers (H\"ofner et al. \cite{HoGAJ03}, Nowotny et al. \cite{NowAH09}).}
\begin{tabular}{l|ll}
\hline
\hline
Model:& P & PM \\
\hline
luminosity $L_\star$ [$L_{\odot}$]&4000&4000\\
mass $M_\star$ [$M_{\odot}$]&0.6&0.6\\
effective temperature $T_\star$ [K]&3500&3500\\
metallicity $[$Fe/H$]$ [dex]&$-0.7$&$-0.7$\\
C/O ratio&0.48&0.48\\
\hline
radius $R_\star$ [$R_{\odot}$]&172&172\\
surface gravity log($g_\star$ [cm/s$^{2}$])&$-0.25$&$-0.25$\\
\hline
piston period $P$ [d]&200&200\\
piston velocity amplitude $\Delta u_{\rm p}$ [km\,s$^{-1}$]&2&3\\
luminosity amplitude parameter $f_{\rm L}$ [1] &4&4\\
bolometric variation $\Delta m_{\rm bol}$ [mag]&0.86&1.36\\
dust mass absorption coefficient $\kappa_{\rm d}^{\rm max}$ [cm$^{2}$\,g$^{-1}$]&--&5\\
\hline
mean mass loss rate $\langle\dot M\rangle$ [$M_{\odot}$yr$^{-1}$]&--&1.9$\cdot$10$^{-7}$\\
mean outflow velocity $\langle u \rangle$ [km\,s$^{-1}$]&--&19.5\\
\hline
\end{tabular}
\label{t:dmaparameters}
\end{center}
\end{table}

\begin{table}
\begin{center}
\caption{Effective temperatures and surface gravities (ranges, stepwidths) of the grid of hydrostatic atmospheric models (COMARCS) used.}
\begin{tabular}{llcl}
\hline
\hline
C/O\,=\,0.48&$T_{\rm eff}$ [K]&2800 $\ldots$ 4000&$\Delta$=50\\
&log($g$ [cm/s$^{2}$])&--0.50 $\ldots$ +0.50&$\Delta$=0.25\\
\hline
C/O\,=\,0.25&$T_{\rm eff}$ [K]&2800 $\ldots$ 4000&$\Delta$=50\\
&log($g$ [cm/s$^{2}$])&--0.25 $\ldots$ +0.75&$\Delta$=0.25\\
\hline
\end{tabular}
\label{t:comarcsparameters}
\end{center}
\end{table}

The guideline for the parameters of our model atmospheres was a typical globular cluster LPV, namely the variable V3 in the well studied cluster 47\,Tuc. This metal poor, short period mira was observed in the past by different authors (cf. Lebzelter \& Wood \cite{LebzW05}, Lebzelter et al. \cite{LWHJF05}+\cite{LPHWB06} and references therein).\footnote{Approximate values found for V3 are: \\ $L$\,=\,4592\,$L_{\odot}$, $T_{\rm eff}$\,=\,3565\,K, P\,=\,192\,d, $\dot M$\,=\,10$^{-7}$--10$^{-6}$ $M_{\odot}$yr$^{-1}$.} The reason for choosing this object (and the corresponding models) was (i) the availability of spectroscopic data,\footnote{To be presented in a follow-up paper.} and (ii) the parameters characterizing the star (comparably high $T_{\rm eff}$, lower metallicity) to reduce the confusing effect of line blending.

For our simulations, we used the dynamic model atmospheres of H\"ofner et al. (\cite{HoGAJ03}). These represent a 
combined and self-consistent solution of hydrodynamics, frequency-dependent radiative transfer\footnote{The 
radiative transfer is solved at 64 frequency points and based on opacity sampling data of molecular opacities.} and 
-- at least in the carbon-rich case -- a detailed time-dependent treatment of dust formation and evolution. Being 
well suited to describing pulsating, mass-losing red giants, these models were able to reproduce observational 
results such
as low-resolution spectra of C-rich stars (Gautschy-Loidl et al. \cite{GaHJH04}), line profile variations in high-
resolution IR spectra (Nowotny et al. \cite{Nowo05a}, \cite{Nowo05b}, \cite{NowAH09}), or interferometric data 
(Paladini et al. \cite{PAHNS09}) reasonably well. The parameters of the atmospheric models (see Table\,
\ref{t:dmaparameters}) were chosen to resemble our reference globular cluster mira. Starting with the same 
hydrostatic initial model, we ended up with two different dynamic models -- P and \mbox{PM --} due to the altered 
piston velocity amplitude $\Delta u_{\rm p}$ of the varying inner boundary (cf. Eqs.\,1 and 2 in Nowotny et al. 
\cite{NowAH09}). While the first one represents a pulsating dust-free atmosphere, we encounter in the second case the 
development of a stellar wind on top of the pulsating photosphere. Plots of moving mass shells and the corresponding 
atmospheric structures during a pulsation cycle for similar models of both types can be found in H\"ofner et al. 
(\cite{HoGAJ03}) or Nowotny et al. (\cite{NowAH09}). We note that in the case of oxygen-rich chemistry the 
physics of dust formation and mass loss of AGB stars is not yet fully understood (e.g. Woitke \cite{Woitk06b}, 
H\"ofner \cite{Hoefn07} \& \cite{Hoefn08}). For M-type models such as those used here\footnote{For the C/O ratios,
 the 
solar value of 0.48 was assumed, dynamic models for other C/O ratios $<$1 are currently under development.} the 
dust-driven wind is therefore induced by a simple parametrised description of the dust opacity first introduced by 
Bowen (\cite{Bowen88}) and described in detail in H\"ofner et al. (\cite{HoGAJ03}; Eq.\,19). We applied a maximum 
dust mass absorption coefficient of $\kappa_{\rm d}^{\rm max}$\,=\,5\,cm$^{2}$\,g$^{-1}$ to generate such an 
artificial wind for model~PM.\footnote{The critical value (i.e. corresponding to $\Gamma$\,=\,$a_{\rm rad}^{\rm 
dust}$/$a_{\rm grav}$\,=\,1) of $\kappa_{\rm d}$ for the chosen set of stellar parameters (Table\,
\ref{t:dmaparameters}) is 2\,cm$^{2}$\,g$^{-1}$ and the chosen $\kappa_{\rm d}^{\rm max}$ may appear to be a little 
high. Still, this is probably not an unrealistic assumption and comparable to what can be estimated for silicates in 
the context of dust-driven winds (see, e.g., H\"ofner \cite{Hoefn08}). Test calculations of another dynamic model with a reduced $\kappa_{\rm d}^{\rm max}$ of 3\,cm$^{2}$\,g$^{-1}$ (but higher piston velocity amplitude)
led to a quite similar behaviour of an outflow strongly driven by radiative pressure. Moreover, not only is the dust 
absorption relevant for the wind, but also the effect of pulsation -- simulated by the piston --
providing kinetic energy input and leading to enhanced densities in the cool outer layers. 
It should indeed be possible to obtain a stellar wind with
an even smaller $\kappa_{\rm d}^{\rm max}$ for a higher $\Delta u_{\rm p}$.}

\begin{figure*}
\sidecaption
\includegraphics[width=12cm,clip]{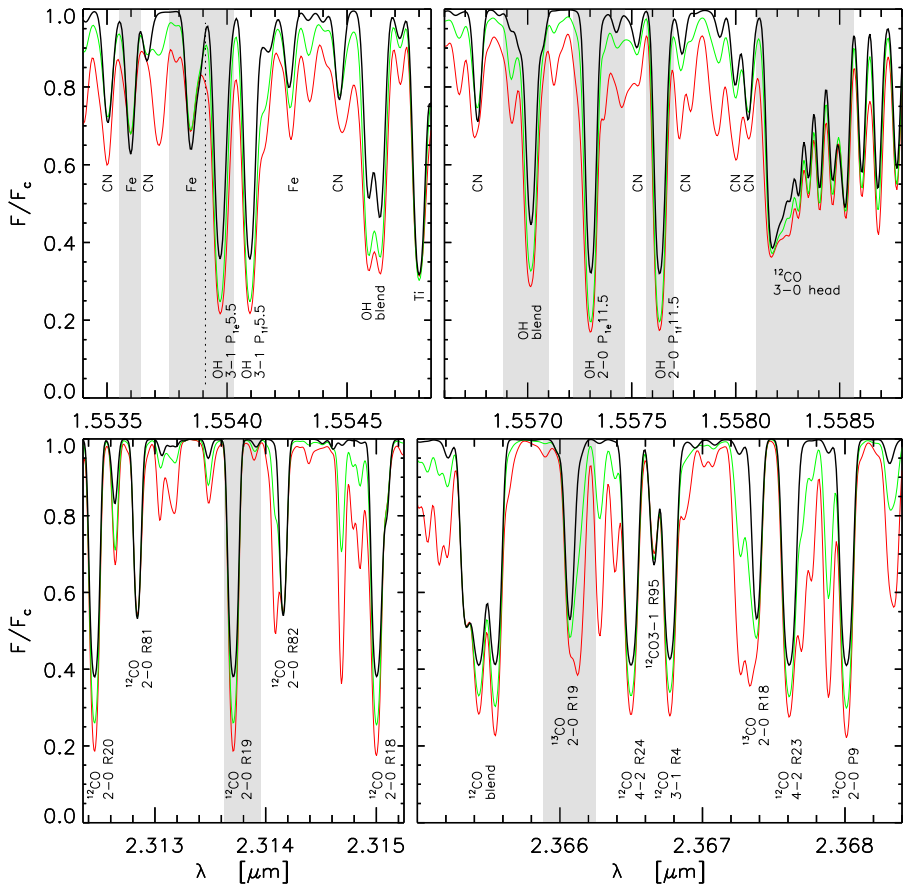}
\caption{Overview of the spectral features studied. Synthetic spectra based on three different COMARCS models 
(hydrostatic, i.e., no velocity effects) with log\,$g$\,=\,0.0 are shown, for $T_{\rm eff}$\,=\,3500\,K (black), 3000\,K (green), and 2800\,K (red), respectively. 
The grey shaded areas indicate the ranges used to measure equivalent widths (cf. Table\,\ref{t:ranges}). The spectra were normalised to the model continuum. Note that for the measurements a local pseudo-continuum was chosen. Some other spectral features are also identified for orientation purposes. Note that for decreasing temperatures spectral features caused mainly by H$_2$O become prominent. For the modelling in this work, we accounted for water with a pseudo-continuous approach (see text in Sect.\,\ref{s:specsynth}) to avoid uncertainties 
caused by line lists of limited precision in wavelength (in contrast to the spectra shown here, where line list data of H$_2$O were utilised).
\newline
\newline
\newline
}
\label{f:linesstudied}
\end{figure*}

Since we wish to study the influence of atmospheric dynamics on spectral lines, we also computed a reference grid of hydrostatic model atmospheres. We thereby followed the approach described in detail by Aringer et al. (\cite{Aringer09}). Being guided again by \mbox{47\,Tuc-V3}, a grid of COMARCS models was calculated with the aim of covering the full colour range of the dynamic models, as illustrated in Figs.\,\ref{f:coloursDMAs} and \ref{f:coloursMAERCSE}. The parameter ranges of the two subgrids with different C/O ratios are given in Table\,\ref{t:comarcsparameters}. Other model parameters were kept constant according to values found in the literature for 47\,Tuc-V3: $M_\star$\,=\,0.6\,$M_{\odot}$ (Lebzelter \& Wood \cite{LebzW05}), $[$Fe/H$]$\,=\,$-0.7$\,dex (Carretta \& Gratton \cite{CarrG97}), $\xi$\,=\,2.5\,km\,s$^{-1}$ (Nowotny et al. \cite{NowAH09}). For the isotopic ratio $^{12}$C/$^{13}$C, we chose a typical AGB value of 30 (cf. Lebzelter et al. \cite{lz08}).

\subsection{Synthetic spectra and photometry}
\label{s:specsynth}

For the spectral synthesis, we followed an approach that had already been
applied very successfully in the past, for details we refer to e.g. Nowotny et al. (\cite{Nowo05a}+\cite{NowAH09}), Aringer et al. (\cite{Aringer02}+\cite{Aringer09}), Lebzelter et al. (\cite{lz08}), or Gautschy-Loidl et al. (\cite{GaHJH04}). 

Based on a given radial atmospheric structure (temperature-pressure) and 
the assumptions of chemical equilibrium as well as conditions of LTE, opacities for all necessary sources 
were calculated by using the COMA code (described in detail by Aringer \cite{Aring00}, Aringer et al. 
\cite{Aringer09}, Lederer \& Aringer \cite{LedeA09}). Element abundances for the spectral synthesis were used  
consistent with the hydrodynamic models. We adopted the values for solar composition from Anders \& 
Grevesse~(\cite{AndeG89}), except for C, N, and O where we took the data from Grevesse \& Sauval~(\cite{GrevS94}).  
To obtain the correct metallicity (cf. Table\,\ref{t:dmaparameters}), the abundances of all elements heavier 
than H and He were then scaled by the same factor of 0.2 (as \mbox{$[$Fe/H$]$\,=\,$-0.7$\,dex} corresponds to 
Z/Z$_{\odot}$\,=\,0.2). The C/O ratios were subsequently adapted by increasing the amount of carbon (without 
changing the rest of the composition). Following the descriptions given in Nowotny et al. (\cite{NowAH09}), Doppler 
profiles were assumed for molecular lines and the corresponding absorption coefficients were computed with the help 
of line lists by Goorvitch \& Chackerian (\cite{GoorC94}) for CO and by Rothman et al. (\cite{RJBCB05}; HITRAN 
database) for OH. For the atomic Fe lines, we followed the method of Lederer \& Aringer (\cite{LedeA09}) and adopted 
full Voigt profiles with line data taken from the VALD database (Kupka et al. \cite{Kupka00}). All sources of 
the continuum opacity listed in Lederer \& Aringer (\cite{LedeA09}) were taken into account. In addition, we included 
absorption of H$_2$O where necessary. Spectral features of water are present across the whole near-IR range and can 
severely influence the spectra for low effective temperatures and mass-losing objects as illustrated in Fig.\,
\ref{f:linesstudied}. However, most of the existing H$_2$O line lists show considerable differences in the positions of lines (Aringer et al. \cite{AriNH08}) when compared with observed spectra. Therefore, we accounted for water in a 
pseudo-continuous way in order not to introduce uncertainties related to the studied line profiles (blending). The water absorption was calculated by using the (low-resolution) OS data from the SCAN data base (J{\o}rgensen \cite{Jorge97}), a detailed description of which can be found in J{\o}rgensen et al. (\cite{JorgJ93}+\cite{JoJSA01}). Spline fits through minimum values of appropriately grouped data points were then applied to determine the 
contribution of H$_2$O to the actual high-resolution wavelength grid. This method ensures that the opacities are not 
overestimated (cf. Fig.\,2.5 in Nowotny \cite{Nowot05}) and infers lower limits to the quasi-continuous 
contributions.

\begin{table}
\centering
\caption{Spectral features investigated. See Fig.\,\ref{f:linesstudied} for an illustration of the feature locations. Column~3 gives an identification for each feature, which is used throughout this paper (containing the corresponding central wavelength). Columns~4+5 list the borders of the spectral ranges used for determining equivalent widths. }
\begin{tabular}{lll|ll}
\hline
Species&transition&ID&$\lambda_{\rm start}$ [{\AA}]&$\lambda_{\rm end}$ [{\AA}]\\
\hline
Fe& &Fe 15536&15535.5&15536.4\\
Fe& &Fe15538&15537.6&15539.1\\
OH&3-1 P$_{\rm 1e}$ 5.5&OH 15540&15539.1&15540.3\\
OH&blend&OH 15570&15568.8&15571.0\\
OH&2-0 P$_{\rm 1e}$ 11.5&OH 15573&15572.2&15574.7\\
OH&2-0 P$_{\rm 1f}$ 11.5&OH 15576&15575.7&15577.0\\
$^{12}$CO&3-0 head&CO 3-0 band head&15581.0&15585.7\\
$^{12}$CO &2-0 R19&$^{12}$CO 23137&23136.3&23139.6\\
$^{13}$CO&2-0 R19&$^{13}$CO 23360&23658.8&23662.5\\
\hline
\noalign{\smallskip}
\end{tabular}
\label{t:ranges}
\end{table}

On the basis of the opacities $\kappa_{\nu}$($\lambda$,r) for all depth and wavelengths points determined in this way, the radiative transfer is solved in spherical geometry. As we wish to investigate the influence of the complex atmospheric velocity fields on spectral lines, a radiative transfer code was applied that takes into account the effects of macroscopic gas velocities. This code follows the algorithm described by Yorke (\cite{Yorke88}) and has
been used successfully in the past (e.g. Nowotny et al. \cite{Nowo05a}+\cite{NowAH09}). Our aim was to study the 
behaviour of the individual line profiles. Thus, we computed synthetic spectra containing the spectral features chosen for our abundance studies (Table\,\ref{t:ranges}) with a high resolution of $R$\,=\,300\,000. Rebinning to $R$\,=\,70\,000 led to spectra with resolutions comparable to observations with state-of-the-art IR spectrographs (FTS, CRIRES, Phoenix). The latter were then used for measuring equivalent widths (Sect.\,\ref{s:measuringwidths}). Sample spectra based on hydrostatic COMARCS models (thus no velocity effects) are shown in Fig.\,\ref{f:linesstudied}.

\begin{figure}
\resizebox{\hsize}{!}{\includegraphics[clip]{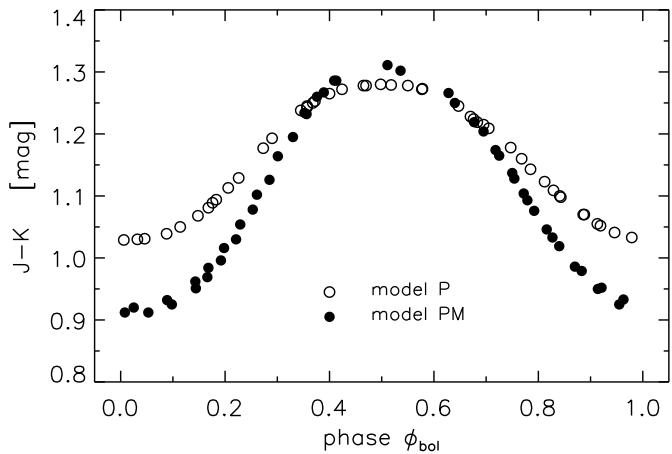}}
\caption{($J$--$K$) colour variation versus bolometric phase for the 
dynamic models with (PM, filled symbols) and without (P, open symbols) wind.}
\label{f:coloursDMAs}
\end{figure}

\begin{figure}
\resizebox{\hsize}{!}{\includegraphics[clip]{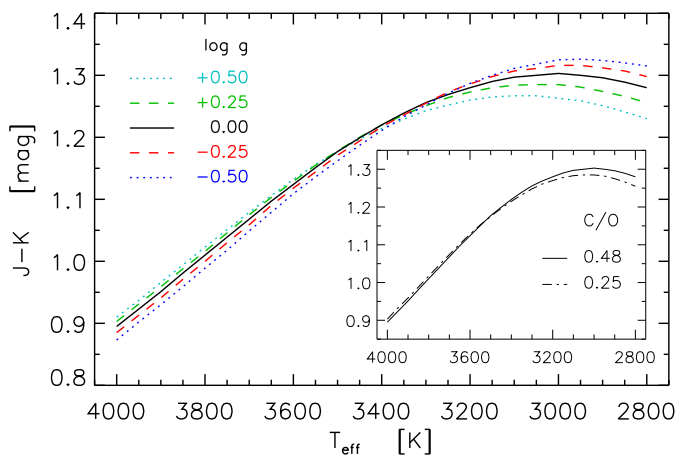}}
\caption{Relation between $T_{\rm eff}$ of the hydrostatic COMARCS model atmospheres and the resulting ($J$--$K$) colours. Different line styles show a variation in log\,$g$ from light grey dotted (log\,$g$=+0.5), light grey dashed (+0.25), solid black (0.0), dark grey dashed ($-$0.25), to dark grey dotted ($-$0.50). In the online colour version, this corresponds to cyan (+0.5), green (+0.25), black (0.0), red ($-$0.25), and blue ($-$0.5), respectively. The insert shows the same relation for log\,$g=$0.0 but two different C/O ratios, illustrating the influence of a change in this parameter.}
\label{f:coloursMAERCSE}
\end{figure}

In addition to these high-resolution spectra containing the line profiles, we also computed low-resolution spectra and synthetic photometry following the approach of Aringer et al. (\cite{Aringer09}). Opacity sampling spectra covering the near-IR with resolutions of R\,=\,10\,000 were calculated on the basis of several atmospheric models. By 
convolving these spectra with filter transmission curves and applying adequate zeropoints, we obtained synthetic 
photometry. The resulting ($J$--$K$) colours (subsequently used to correlate different models) for the dynamic model 
atmospheres during the pulsation cycle and for the grid of hydrostatic COMARCS models, are shown in Figs.\,
\ref{f:coloursDMAs} and \ref{f:coloursMAERCSE}, respectively.

\subsection{Measurement of equivalent widths}
\label{s:measuringwidths}

Elemental abundances in the atmospheres of stars can be derived from the intensity of spectral features since the number of absorbers will directly affect the depth of a line (except for 
saturated lines) apart from effects of temperature and electron density. Since we wish to investigate the effect of atmospheric dynamics on the derived abundances, we have 
to look for a quantity describing the line intensity. The standard concept in the context of abundance measurement is 
the equivalent width of a line, which is related to the abundance of the parent element by the curve of growth (e.g.
\,Castelli \& Hack \cite{CH90}). The spectra of cool giants with their severe blending of adjacent molecular lines 
would certainly favour the application of a full spectral comparison to study the abundances of the elements on the 
surface (see e.g.\,Lebzelter et al.\,\cite{lz08}). However, as we see below, stellar pulsation leads to considerable 
modifications of the line profile with the appearance of emission or doubled absorption components (cf. Nowotny et 
al.\,\cite{Nowo05a}, \cite{NowAH09}). Therefore, a complete fit of a spectrum of a mira-like variable in a particular 
spectral range  with hydrostatic model spectra will fail anyway (at least during a good portion of the pulsation 
cycle). This makes a comparison of hydrostatic and dynamic models difficult. With the help of the equivalent width, we can at least measure the amount by which the total line intensity is changed by pulsation.

We obtained equivalent widths for a selection of spectral features of various species. For this, we first imported the synthetic spectra into IRAF. To allow for an automatic and objective measurement, we removed all velocity shifts 
from the dynamical spectra by shifting the investigated features to zero velocity (while retaining the Doppler-
modified line profiles). For this shift, we used the line core as a reference point, except in cases of line doubling,
where we used instead a mean value between the two components. For the shifted spectra, we afterwards verified that
the complete feature falls within our range of measurement and corrected the shift if necessary. All spectra were normalized to the local pseudo-continuum instead of the model continuum\footnote{In contrast to the pseudo-continuum approach usually applied by observational studies, modelling offers the possibility to normalize the synthetic spectra relative to a computation for which only the continuous opacity was taken into account ($F$/$F_{\rm cont}$; cf. Fig.\,\ref{f:linesstudied}).} to resemble an observer's approach as realistically as possible. The pseudo-continuum was defined by placing a straight line through the highest points in the spectrum excluding obvious emission peaks (as present in the case of the dynamic models). The equivalent widths were then measured using the IRAF tasks {\tt splot} and {\tt bplot}.

For our study, we selected lines in the $H$- and $K$-band. These wavelength ranges have played a key role in the study of photospheres of AGB stars since the development of high resolution near-infared spectrographs. Within this wavelength range, we chose a combination of strong and weak lines, and atomic and molecular features including a molecular band head for our analysis. 
Many of the selected lines had already been used in the abundance analysis of red giants (e.g. Smith et al. \cite{Smith02}, Lebzelter et al. \cite{lz08}). For each feature we measured the same area around the central wavelength in all spectra. The borders are listed in Table\,\ref{t:ranges}. Examples illustrating the measured range are shown in Fig.\,\ref{f:linesstudied}.

A few uncertainties in this kind of measurement have to be considered. First, variable line asymmetries as well as a 
limited number of points in the pseudo-continuum can introduce small inconsistencies in the handling from one 
spectrum to another. We ran some tests and found possible deviations below 2\% of the measured equivalent width. A 
second and more severe problem is blending, although we tried to avoid strongly blended lines in our selection of 
spectral features. The problem emerges mainly for cool stars ($T_{\rm eff}<$3000\,K) with the appearance of H$_{2}$O 
and CN lines in the spectrum. We return to this point below. For the dynamic models, the occurrence of line 
doubling and emission components is a problem. The former can lead to significant broadening of the feature 
requiring us to measure a wider range about the central wavelength. This, however, leads to an inclusion of any 
(typically blueshifted) emission component reducing the equivalent width of the feature at other phases. When 
measuring a line in an observed stellar spectrum, one may deliberately exclude this emission component provided the 
spectral resolution is high enough. This would lead to qualitative and phase-dependent differences between our model results and observations. Emission components close to the line core do not require special consideration since any measurement in observed spectra would not be able to distinguish them either.

\section{Dependence of hydrostatic line sensitivities on stellar parameters} 
\label{s:hydro}

We describe the dependency of the intensity of various spectral features on (i) effective 
temperature ($T_{\rm eff}$), (ii) surface gravity (log\,$g$), (iii) C/O ratio, and (iv) $^{12}$C/$^{13}$C 
ratio as derived from synthetic spectra based on hydrostatic models. The results are grouped by the element or 
molecule producing the feature.

As can be seen from Fig.\,\ref{f:coloursMAERCSE}, the roughly linear relation between $T_{\rm eff}$ and ($J$--$K$) stops at a temperature of 3200\,K. For lower temperatures, the near-infrared colour becomes almost constant and even shows a slight linear trend towards lower values. This phenomenon has also been found for carbon rich stars by Aringer et al. 
(\cite{Aringer09}) and is caused by the behaviour of different molecular features in the region of the filters. For 
the same reason -- but of course due to different molecular species (mainly H$_2$O) -- a similar trend is seen in the 
oxygen-rich models. However, AGB stars in this low-temperature range are subject to mass-loss processes and the 
effect induced on the colour by the subsequent circumstellar reddening may be much larger than the one caused by the 
aforementioned behaviour of molecular features. This is especially true for C-rich stars and high mass-loss rates (Aringer et al. \cite{Aringer09}).

As shown below, the intensities of the spectral features also continue to increase below 3200\,K (except for the iron 
lines, which decrease in strength with decreasing $T_{\rm eff}$). This results in a turnaround in the equivalent width curves at the high ($J$--$K$) values. We decided to plot the relations versus the near infared colour for two reasons. First, the ($J$--$K$) colour is a quantity that is 
directly measurable. It is thus easier to compare the modelling results with observations (see below). Second, the comparison with dynamic models (performed in the next section) requires the use of a global parameter such as 
the colour, because the concept of an effective temperature is questionable for extended dynamical structures.

\subsection{Fe lines}

\begin{figure}
\resizebox{\hsize}{!}{\includegraphics[clip]{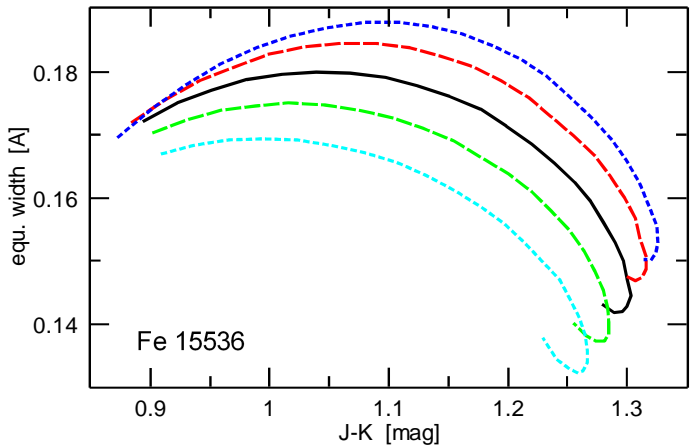}}
\caption{Equivalent widths in [{\AA}] of the Fe line at 15\,536\,{\AA} measured from synthetic spectra based on hydrostatic model atmospheres against the corresponding ($J$--$K$) values. Same line style coding for various log\,$g$ values as in Fig.\,\ref{f:coloursMAERCSE}. Details of the range in wavelength used around the line centre are listed in Table\,\ref{t:ranges} and illustrated in Fig.\,\ref{f:linesstudied}.}
\label{f:Fe15536}
\end{figure}

Two iron lines were selected for the analysis to investigate the behaviour of lines typically used to 
derive stellar metallicities (e.g. Smith et al.\,\cite{Smith02}). The two lines originate in the same excitation level (5.642 eV). Accordingly, both lines show a similar trend with $T_{\rm eff}$ 
and log\,$g$, there being a clear dependence of the equivalent widths on both parameters as illustrated for one
of the lines in Fig.\,\ref{f:Fe15536}. The line at 15\,536\,{\AA} is only mildly affected by blending, although a CN 
feature becomes strong for the lowest temperatures studied. The second Fe line close to 15\,538\,{\AA},
which behaves qualitatively identically, is blended with a nearby CN line producing the asymmetric line profile
clearly visble in Fig.\,\ref{f:linesstudied}. A neighbouring OH line mildly
affects the red wing of the line profile. As can be seen from Fig.\,\ref{f:linesstudied}, a large fraction of the 
observed decrease in the line strengths at low temperatures is caused by a depression of the overall continuum.

\subsection{OH lines}

Lines of the OH molecule produce very prominent features in the $H$-band spectra of cool giants. We selected two lines from the 2-0 transition and one line from the 3-1 transition for our study. In addition, we included a blend of OH lines across the same wavelength region in our analysis. A CN line is located in the blue wing of this blend. The blue end for measuring the equivalent width was defined to include the complete blend also in the dynamical case, but this means that most of the CN line is also covered (see Fig.\,\ref{f:linesstudied}).

\begin{figure}
\resizebox{\hsize}{!}{\includegraphics[clip]{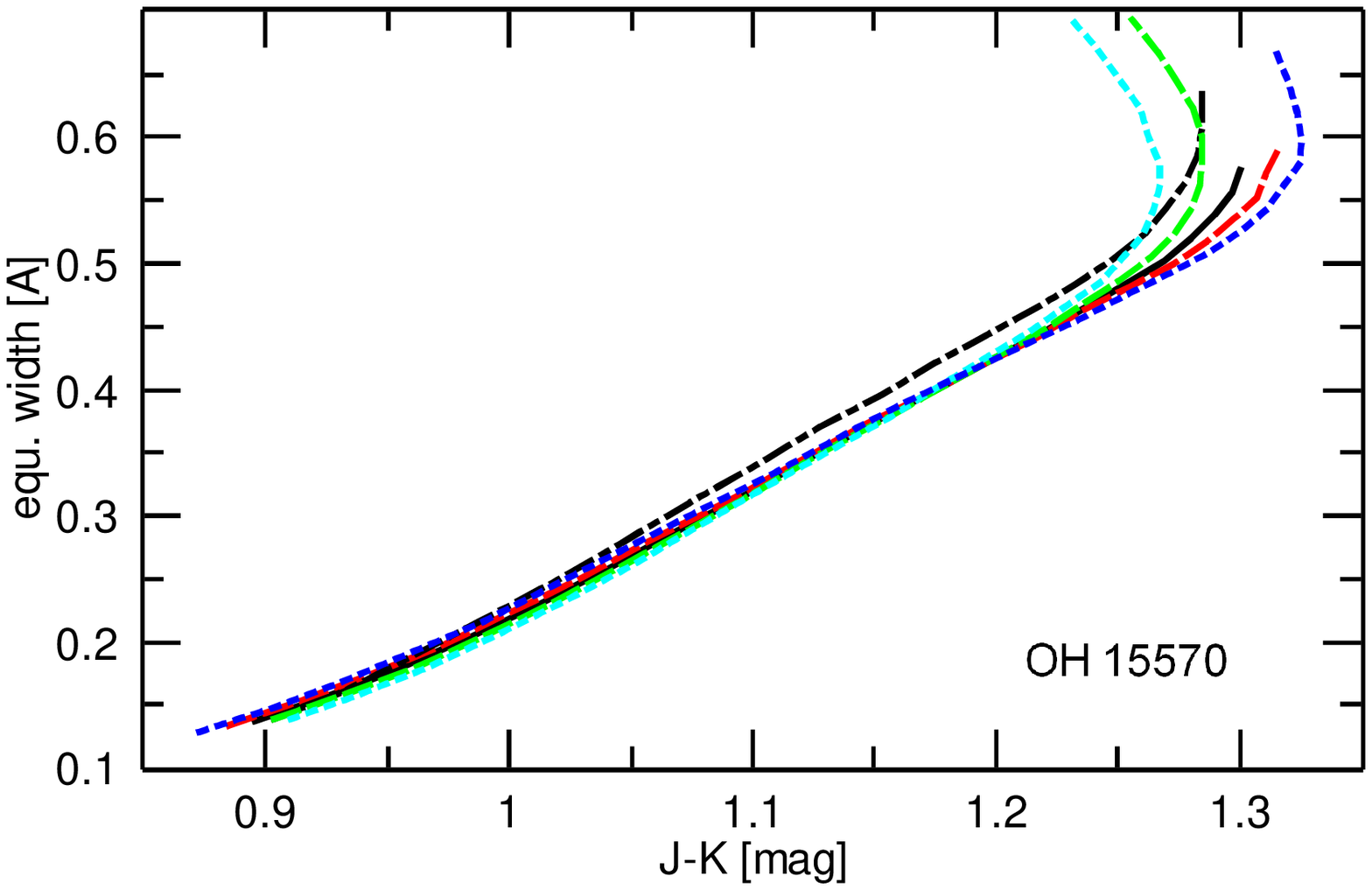}}
\caption{Equivalent widths in [{\AA}] of the OH blend at 15\,570\,{\AA} measured from synthetic spectra based on hydrostatic model atmospheres against the corresponding ($J$--$K$) values. Same line style coding for various log\,$g$ values as in Fig.\,\ref{f:coloursMAERCSE}. The dash-dotted line indicates the relation for C/O$=$0.25 (standard value: 0.48).}
\label{f:OH15570}
\end{figure}

\begin{figure}
\resizebox{\hsize}{!}{\includegraphics[clip]{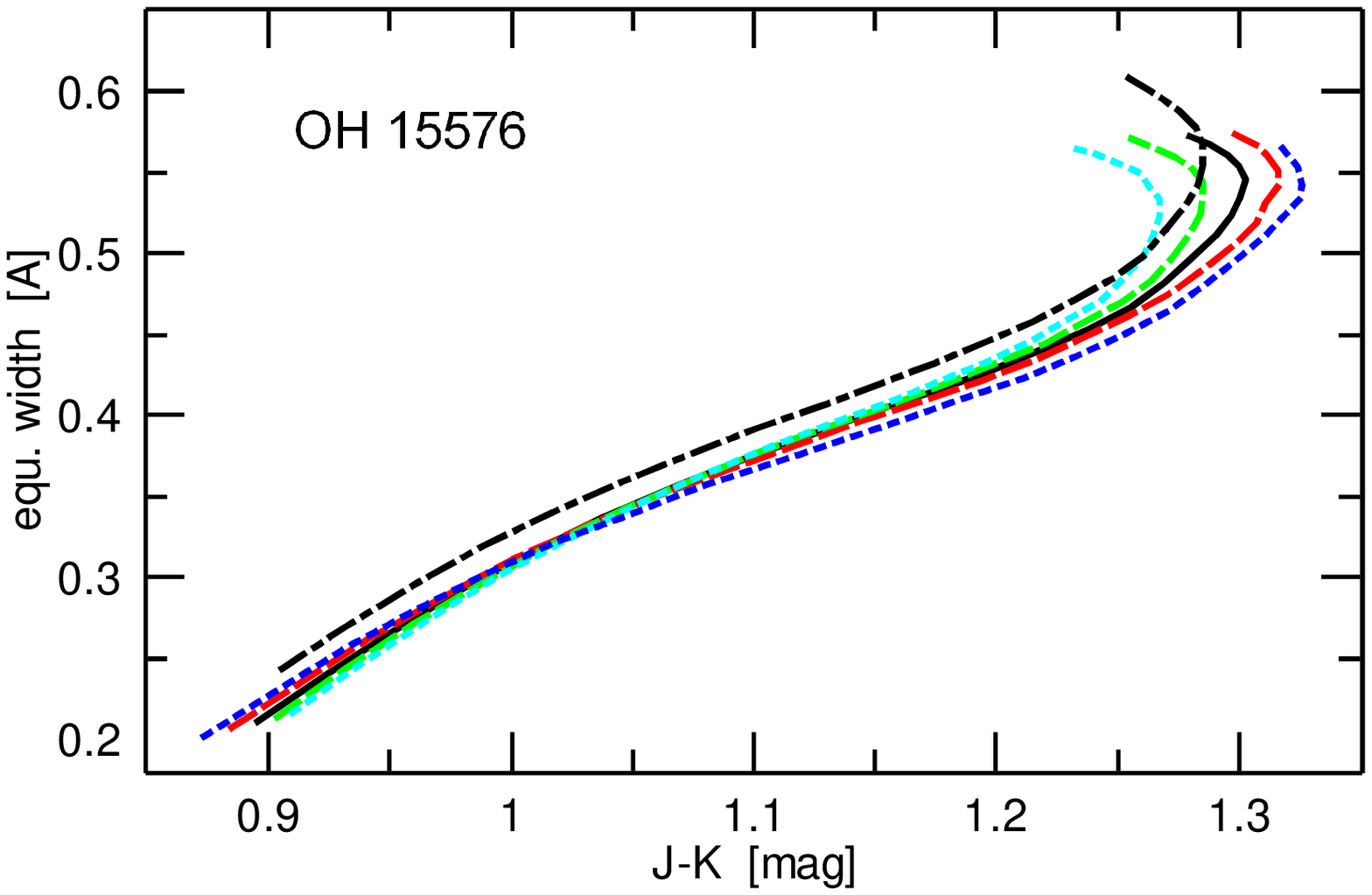}}
\caption{Same as Fig.\,\ref{f:OH15570} for the OH line at 15\,576\,{\AA}.}
\label{f:OH15576}
\end{figure}

All these lines behave identically. They show a strong dependency on ($J$--$K$), i.e. $T_{\rm eff}$, while the change 
in the line intensity with log\,$g$ is comparably small. This is shown for two of the lines in Figs.\,\ref{f:OH15570} 
and \ref{f:OH15576}. The OH blend (Fig.\,\ref{f:OH15570}) including also a CN line does not behave differently, the 
feature being dominated by OH. We also derived equivalent widths for a model with log\,$g$\,=\,0.0 and C/O\,=\,0.25 to 
study the effect of a change in the C/O ratio. The other parameters were left unchanged. The result is plotted as a 
dash-dotted line in the figures. The effect is visible for the OH 2-0 line at 15\,576\,{\AA}, but it is rather small. 
In this context, we note that for our modelling -- simulating the change caused by third dredge up -- the 
C/O ratio is changed by increasing or decreasing the carbon abundance, while the oxygen abundance remains constant.

\begin{figure}
\resizebox{\hsize}{!}{\includegraphics[clip]{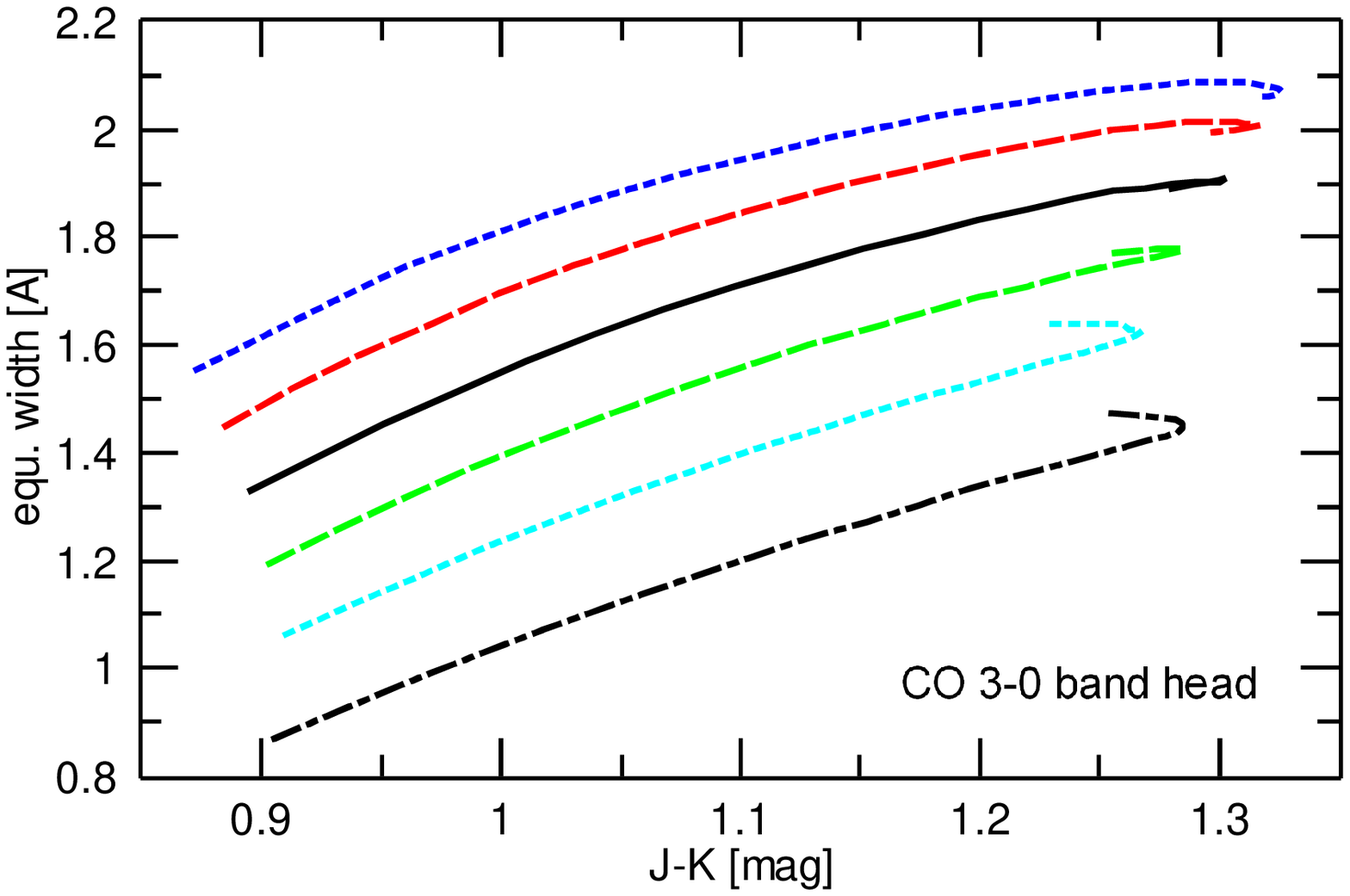}}
\caption{Equivalent widths in [{\AA}] of the $^{12}$CO 3-0 band head (measured between 15\,581\,{\AA} and 
15\,585.7\,{\AA}) measured from synthetic spectra based on hydrostatic model atmospheres against the corresponding 
($J$--$K$) values. See Fig.\,\ref{f:linesstudied} for the wavelength range covered. Same line style coding for 
various log\,$g$ values as in Fig.\,\ref{f:coloursMAERCSE}. The dash-dotted line indicates the relation for 
C/O$=$0.25 (standard value: 0.48).}
\label{f:CO_band}
\end{figure}

\begin{figure}
\resizebox{\hsize}{!}{\includegraphics[clip]{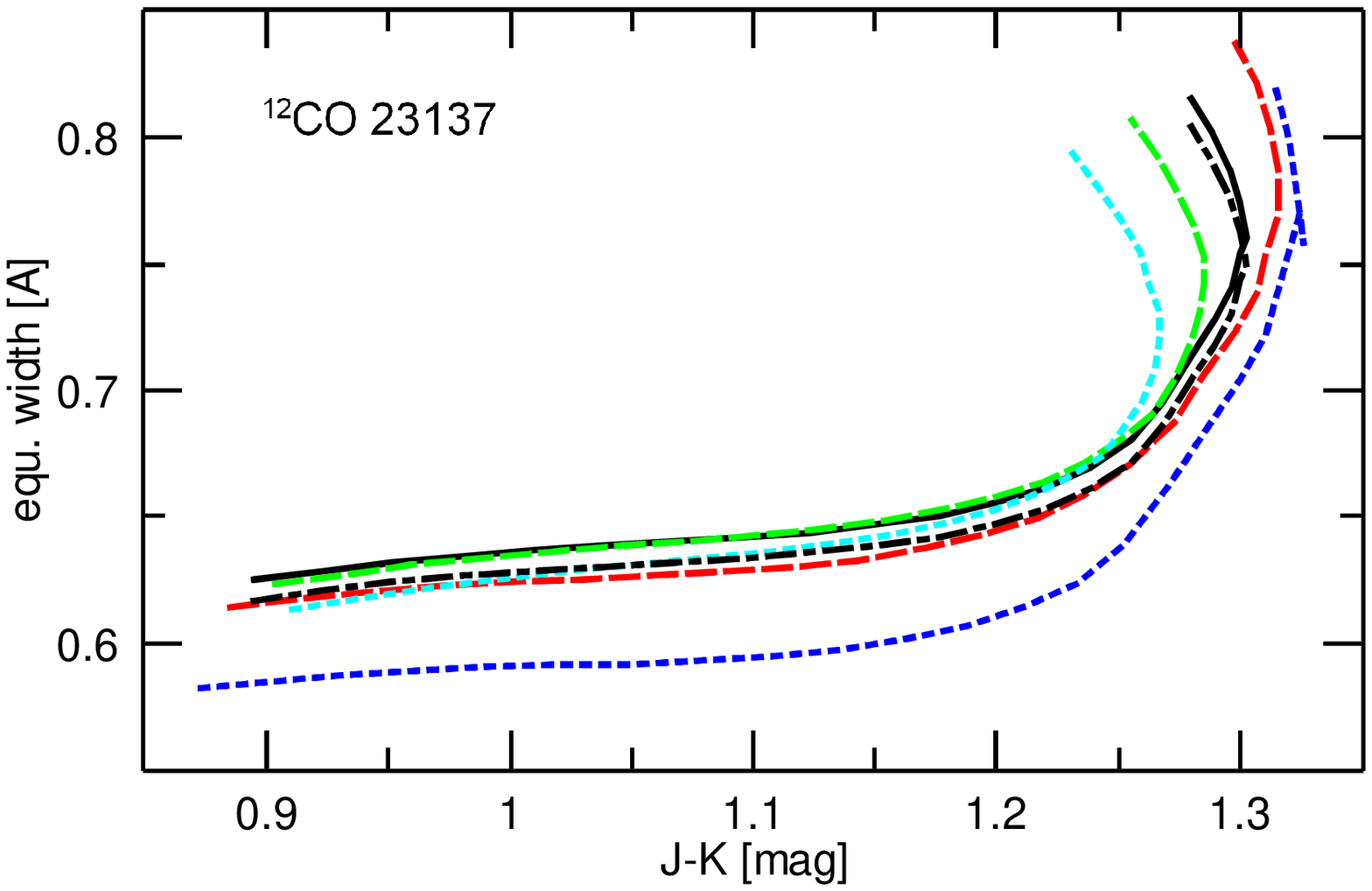}}
\caption{Equivalent widths in [{\AA}] of the $^{12}$CO line at 23\,137\,{\AA} measured from synthetic spectra based on hydrostatic model atmospheres versus the corresponding ($J$--$K$) values. Same line style coding for various log\,$g$ values as in Fig.\,\ref{f:coloursMAERCSE}. The dash-dotted line indicates the relation for $^{12}$C/$^{13}$C$=$10 (standard value: 30).}
\label{f:COR19}
\end{figure}

\begin{figure}
\resizebox{\hsize}{!}{\includegraphics[clip]{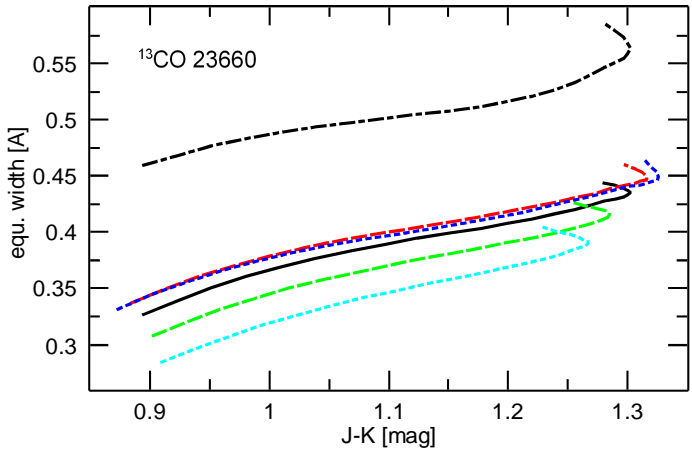}}
\caption{Same as Fig.\,\ref{f:COR19} for the $^{13}$CO line at 23\,660\,{\AA}.}
\label{f:13CO}
\end{figure}

\subsection{CO features}
 
Features of CO with its well expressed band structure can be easily identified in the $H$- and $K$-band spectra of stars of spectral type K or later. They are frequently used to derive stellar parameters and abundances or to study the atmospheric dynamics (e.g. Hinkle et al.\,\cite{HHR82}, Lebzelter \cite{Lebzelter99}). The three CO features selected here represent different spectral signatures of this molecule: the (3-0) band head of $^{12}$CO located in the $H$-band; a strong single $^{12}$CO line in the $K$-band; and finally a line from $^{13}$CO in the same wavelength range.

As in all other cases, the equivalent widths of the three features were derived for various $T_{\rm eff}$ and log\,$g$ (Figs.\,\ref{f:CO_band} to \ref{f:13CO}). In addition, the CO band head was also measured for a model with log\,$g=$0.0 and reduced C/O$=$0.25. To investigate the effect of a change in the carbon isotopic ratio, we also calculated synthetic spectra with $^{12}$C\,/\,$^{13}$C$=$10 instead of 30 (log\,$g=$0.0). 

The strength of the CO band head (Fig.\,\ref{f:CO_band}) depends on all three parameters $T_{\rm eff}$, log\,$g$, and 
C/O. As one would expect from simple considerations concerning molecule formation, for the star investigated
here the intensity of the C/O bands increases with the C/O ratio.
Quite a different behaviour can be seen for the $^{12}$CO line at 23137\,{\AA}. The equivalent width is almost 
constant for a wide range of ($J$--$K$), but below 3400\,K a strong dependency on $T_{\rm eff}$ is observed. We
note that, in contrast to the situation around 1.5\,$\mu$m, 
the line strengths of the features at 2.3\,$\mu$m are not affected 
by a change in the continuum level. The $^{12}$CO line investigated is almost independent of log\,$g$. An exception 
is the model with the lowest surface gravity, which seems to be offset from the other models. The reason for 
that behaviour is unclear, but a weakening of the lines due to sphericity effects occurring in
very extended atmospheres is a likely explanation (Aringer et al.\,\cite{Aringer09}). Finally, 
the $^{13}$CO line again shows a trend of increasing line strenghts with $T_{\rm eff}$ and a mild dependency on the 
log\,$g$ value (especially for log\,$g<$0.0, see Fig.\,\ref{f:13CO}). The change with $T_{\rm eff}$ seems to be 
intensified by a neighbouring line of CN, which appears at the lower end of the temperature range studied. A
change in the carbon isotopic ratio has a strong impact on the intensity of the studied  $^{13}$CO line.

\section{Comparing dynamical and hydrostatic line intensities}
\label{s:dynvsstat}

We summarize our findings for hydrostatic models as follows:
\begin{itemize}

\item Most features are sensitive to changes in $T_{\rm eff}$. The only exception is the $^{12}$CO line at 23\,137\,{\AA}, which becomes temperature sensitive for low temperatures only. The two studied lines of Fe are 
also less sensitive to $T_{\rm eff}$ changes for higher temperatures.

\item The dependency on $T_{\rm eff}$ is a combination of a change in both the line intensity and the level of the pseudo-continuum due to H$_2$O. 

\item At the lowest temperatures, we observed CN line blending for most of the features studied.

\item A change in surface gravity has on average a smaller influence on the feature strengths. The most pronounced changes in this parameter were found for the CO 3-0 band head and the iron lines.

\end{itemize}

With this study of the hydrostatic case we defined a reference grid with which to compare 
the dynamic models. We again discuss the effects of pulsation on a line by line basis.

\subsection{Fe lines}
 
\begin{figure}
\resizebox{\hsize}{!}{\includegraphics[clip]{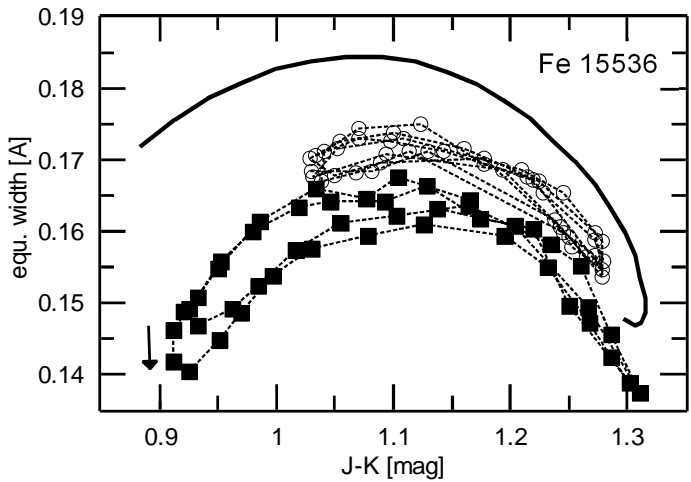}}
\caption{Change in the intensity of the Fe line at 15\,536\,{\AA} over the calculated pulsation cycles versus ($J$--$K$) for two dynamic models. The model with mass loss (PM) is shown with solid symbols, while the model without mass loss (P) is marked with open symbols. Data points are connected to illustrate the path of the change over the pulsation cycle. Two cycles are plotted for model~PM and four for model~P. The small arrow indicates the direction of the path and starts at the maximum bolometric phase ($\phi_{bol}=$0). The solid line shows the hydrostatic model for log\,$g=-$0.25 for comparison (see Sect.\,\ref{s:hydro}). Compare with Fig.\,\ref{f:Fe15536}
for various values of log\,$g$.}
\label{f:Fe15536dyn}
\end{figure}

\begin{figure}
\resizebox{\hsize}{!}{\includegraphics[clip]{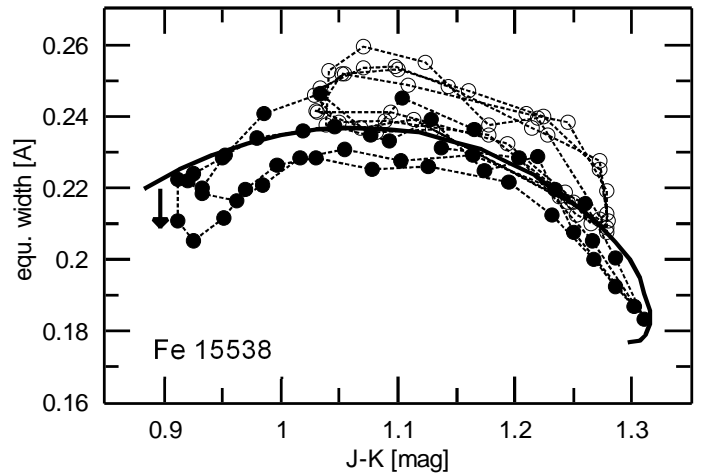}}
\caption{Same as Fig.\,\ref{f:Fe15536dyn} for the Fe line at 15\,538\,{\AA}.}
\label{f:Fe15538dyn}
\end{figure}

We first note here the overall behaviour of dynamic models in the ($J$--$K$) versus equivalent width plane, 
which is already clearly visible for the Fe lines in Figs.\,\ref{f:Fe15536dyn} and \ref{f:Fe15538dyn}: for a given ($J$--$K$) value, the line intensity is different for the falling and rising branch of the light curve. As a result, 
the model values loop through the diagram. This loop is more pronounced in the case of the model with mass loss (PM) than for the model without mass loss (P). Small changes are visible between the pulsation cycles calculated, but the general behaviour remains the same.

In the dynamcial case, the ($J$--$K$)\,--\,equivalent width relations for the iron lines show a similar arc-like shape as in the hydrostatic case. The maximum line intensity is reached at neither maximum nor minimum light. 
However, the two iron lines clearly differ in their behaviour. Fe\,15\,538\,{\AA} seems to vary in a way similar to
that in the hydrostatic models. The variation in line intensity over the pulsation cycle for model~PM may be described by a combination of a hydrostatic model of log\,$g=$0.0 for the first half of the light curve and a hydrostatic model of somewhat lower surface gravity for the second half of the light variation. Model P is offset to somewhat higher line strengths.

The Fe\,15536\,{\AA} line (Fig.\,\ref{f:Fe15536dyn}) shows a trend clearly different from the other iron line and from the hydrostatic reference case. Here the line strenghts are weaker than in the COMARCS model and would correspond to a higher surface gravity. In addition, the shape of the relation is different. Model~P again exhibits
stronger lines at a given $(J-K)$ than model~PM. For $(J-K) >$1.2, model~P approaches the hydrostatic relation.

\subsection{OH lines}

All four OH lines investigated exhibit similar behaviour over the pulsation cycle. As for the iron lines, 
the model produces a loop in the diagram over one pulsation cycle. The dynamic model~P again exhibits 
stronger lines than model~PM. The difference between the two models changes from line to line. 

\begin{figure}
\resizebox{\hsize}{!}{\includegraphics[clip]{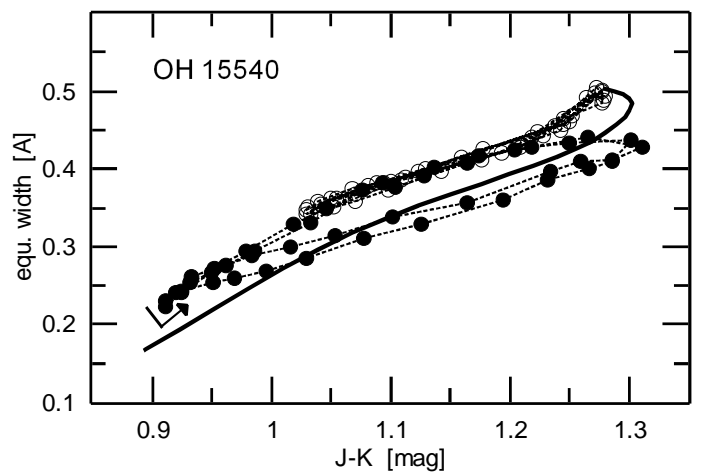}}
\caption{Same as Fig.\,\ref{f:Fe15536dyn} for the OH line at 15540\,{\AA}.}
\label{f:OH15540dyn}
\end{figure}

\begin{figure}
\resizebox{\hsize}{!}{\includegraphics[clip]{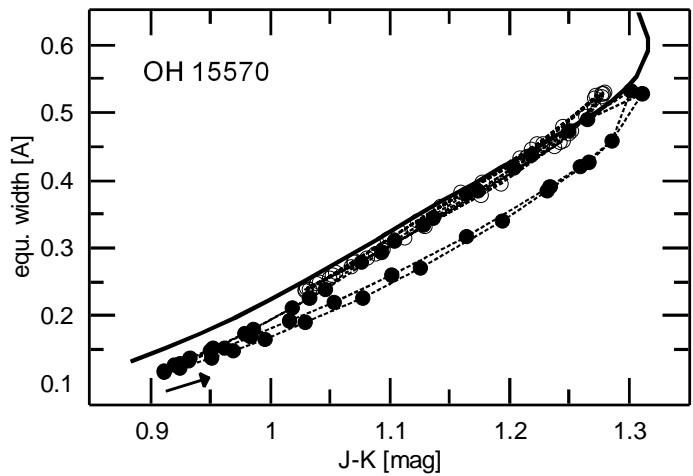}}
\caption{Same as Fig.\,\ref{f:Fe15536dyn} for the OH blend at 15570\,{\AA}. Compare with Fig.\,\ref{f:OH15570}.}
\label{f:OH15570dyn}
\end{figure}

\begin{figure}
\resizebox{\hsize}{!}{\includegraphics[clip]{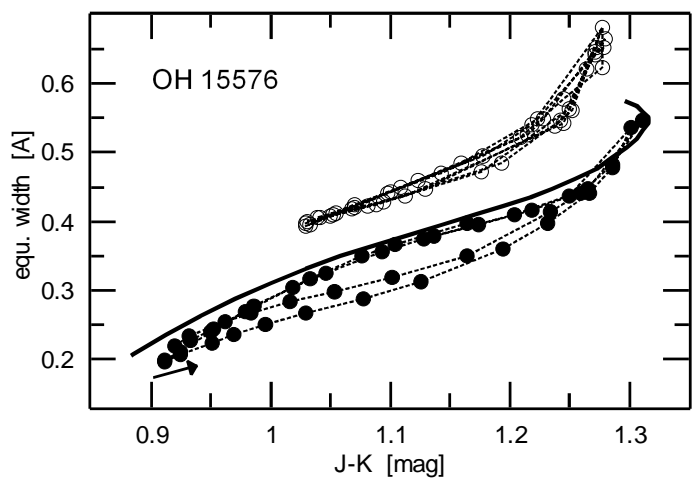}}
\caption{Same as Fig.\,\ref{f:Fe15536dyn} for the OH line at 15576\,{\AA}. Compare with Fig.\,\ref{f:OH15576}.}
\label{f:OH15576dyn}
\end{figure}

Both dynamic models have similar dependences on \mbox{($J$--$K$)}, i.e. $T_{\rm eff}$, as our hydrostatic reference 
models (Figs.\,\ref{f:OH15540dyn} to \ref{f:OH15576dyn}). During half of the pulsation cycle, we can see an offset between model~PM and the hydrostatic case. As the hydrostatic line intensity is almost independent of the surface 
gravity (cf. Figs.\,\ref{f:OH15570} and \ref{f:OH15576}), this offset cannot be compensated for 
by choosing a different log\,$g$ value as in the case of the iron lines. To bring the dynamic model in alignment with a series of hydrostatic models, one would be inveigled to change the C/O ratio. However, the OH line at 
15\,540\,{\AA} (Fig.\,\ref{f:OH15540dyn}) shows a deviation from the hydrostatic case towards the opposite side of 
the other OH lines, therefore requiring an inverse change of C/O.

The different slope seems to be a result of a complex mixture of various components of the investigated lines, which are difficult or even impossible to distinguish in the measurement. One obvious feature is the appearance of an emission component between $\phi_{bol}=$\,0.26 and 0.41 (corresponding to the lower branch of the loop between ($J$--$K$)\,=\,1.1 and 1.28; see Fig.\,\ref{f:OH15576dyn}) 
on the blue wing of the OH 15573 and OH 15576\,{\AA} lines. This emission component is not excluded from the 
measurement of the equivalent width and may be responsible for at least part of the depression of the line 
intensity measured.

\subsection{CO features}
 
\begin{figure}
\resizebox{\hsize}{!}{\includegraphics[clip]{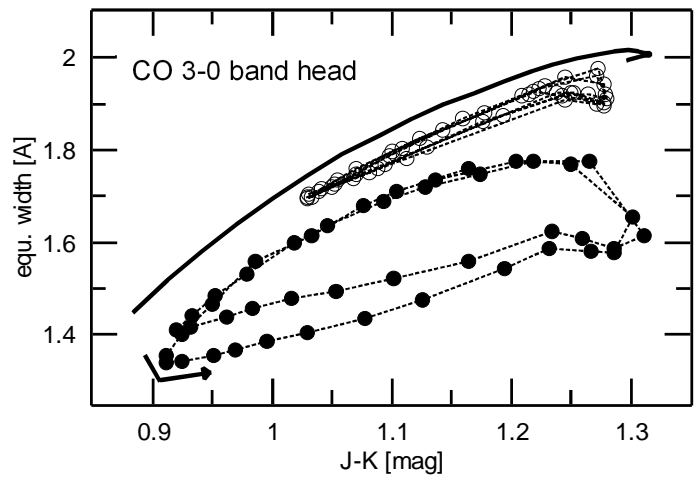}}
\caption{Same as Fig.\,\ref{f:Fe15536dyn} for the $^{12}$CO 3-0 band head. Compare with Fig.\,\ref{f:CO_band}.}
\label{f:COband_dyn}
\end{figure}

The most remarkable loops in the diagram EW versus ($J$--$K$) corresponding to model~PM are found for the features of 
CO. For the CO band head (Fig.\,\ref{f:COband_dyn}), model~P (no outflow) and part of the pulsation cycle of model~PM show a very similar relation of band strength with ($J$--$K$) as the hydrostatic models. 
Owing to the high sensitivity of the CO band head to the surface gravity 
(note the different scale on the y-axis in Fig.\,\ref{f:CO_band}), the whole loop made by model~PM can be
reproduced by a change in log\,$g$ of 0.25. A modification of the C/O ratio from 0.48 to 0.25 has an about three times larger effect on the band strength than the changes during the pulsation cycle.

\begin{figure}
\resizebox{\hsize}{!}{\includegraphics[clip]{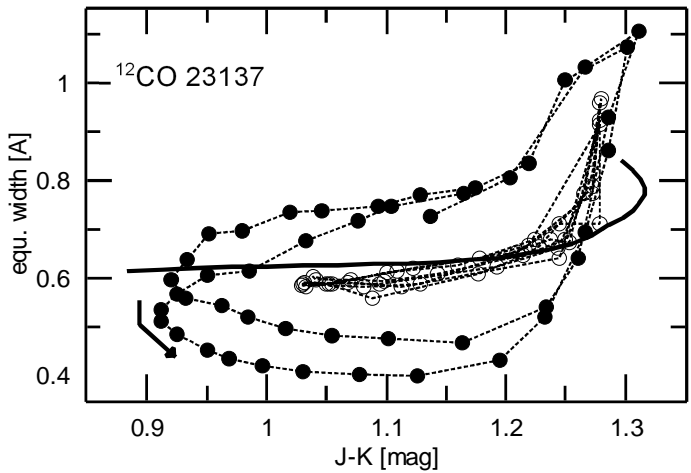}}
\caption{Same as Fig.\,\ref{f:Fe15536dyn} for the $^{12}$CO line at 23137\,{\AA}. Compare with Fig.\,\ref{f:COR19}.}
\label{f:COR19dyn}
\end{figure}

\begin{figure}
\resizebox{\hsize}{!}{\includegraphics[clip]{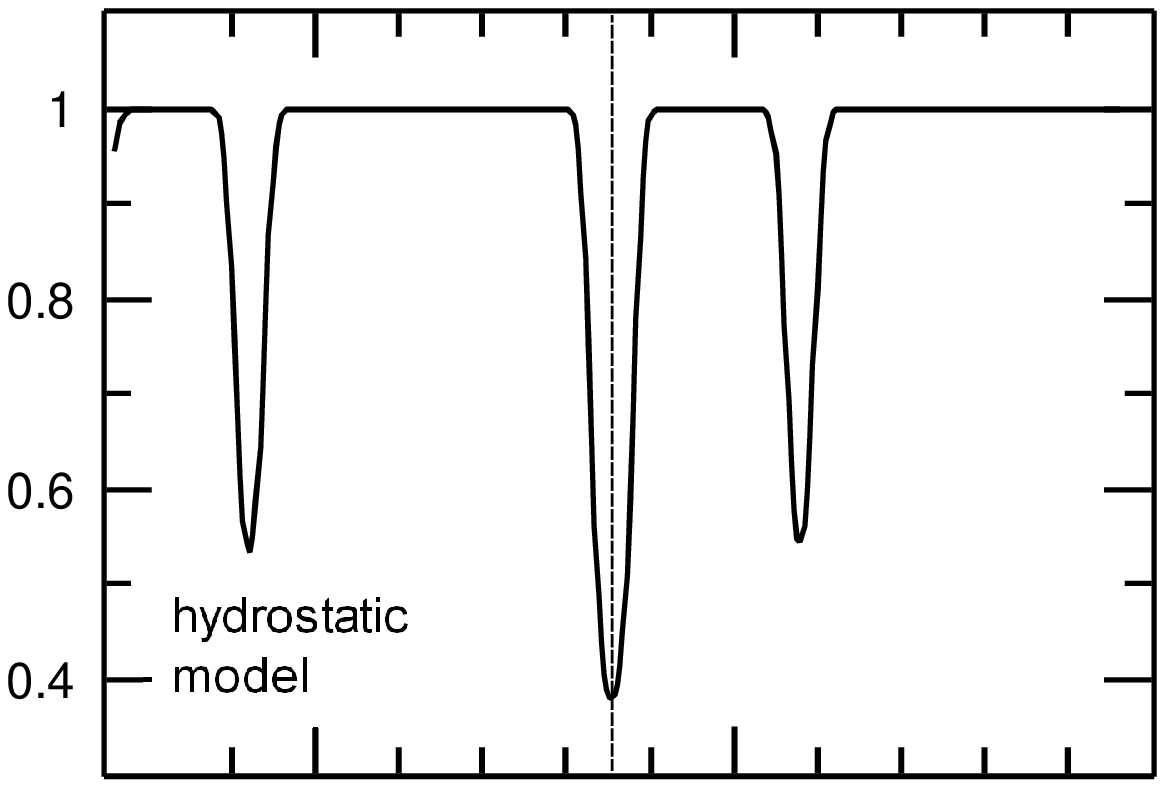}}
\resizebox{\hsize}{!}{\includegraphics[clip]{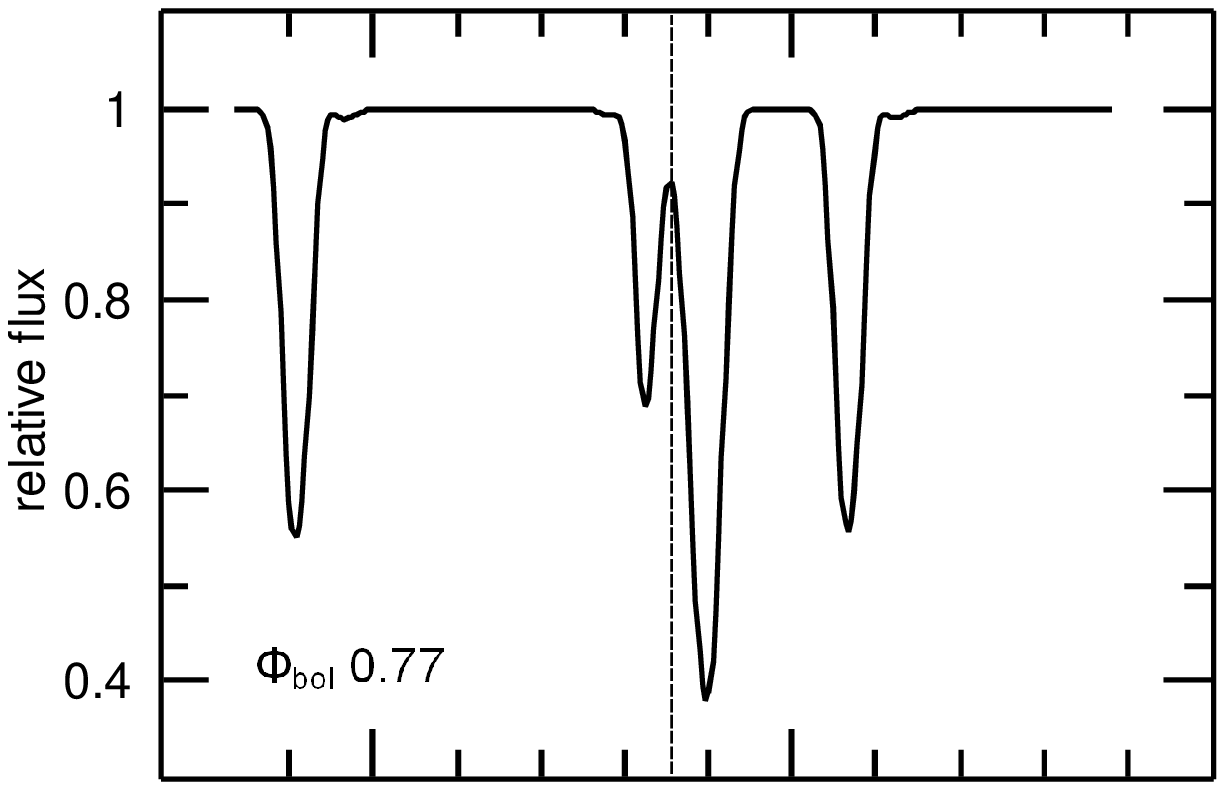}}
\resizebox{\hsize}{!}{\includegraphics[clip]{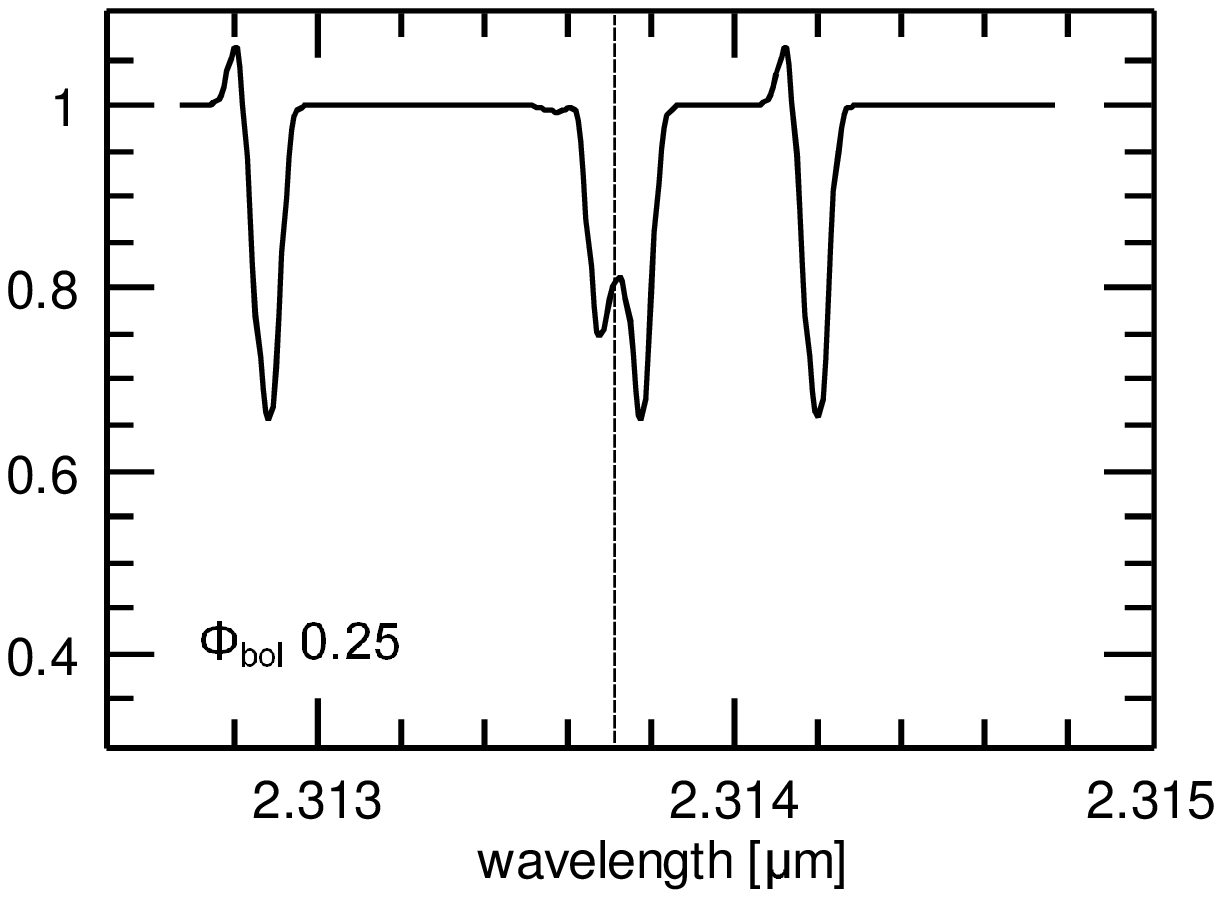}}
\caption{Comparison of the line profile of $^{12}$CO 23\,137 in a  hydrostatic model ($T_{\rm eff}=$3650\,K, log\,$g=$0.0) and two phases of the dynamic model~PM. All three show a very similar ($J$--$K$) value close to 1.1.}
\label{f:lineprofiles}
\end{figure}

For the two individual CO lines studied in the $K$-band, the situation is completely different, especially for model~PM (with outflow). For the $^{12}$CO line at 23\,137\,{\AA}, we find only very short parts of the pulsation cycle where the derived line intensity is in a similar range as that found for the hydrostatic models. This deviation from the hydrostatic case is caused by the occurrence of line doubling over most of the pulsation cycle. However, 
this does not lead to a simple broadening. As can be seen from Fig.\,\ref{f:COR19dyn}, the line intensity at some phases is also weaker than in the hydrostatic case. 
This is caused by an emission component that is visible on the blue side of the line profile for almost half the light cycle. 
For illustration, we compare in Fig.\,\ref{f:lineprofiles} the corresponding line profile of this CO line derived 
from a hydrostatic model and from two phases of model~PM at similar ($J$--$K$). The two chosen phases represent the 
two branches of the bolometric light curve.

Model P resembles the hydrostatic case more closely. Line doubling is observed less often, but again a blue-shifted 
emission component can be seen. Here the differences between hydrostatic and dynamic models are greatest at minimum light.

Line profile variations such as those havee frequently been reported in the literature 
(e.g. Hinkle et al. \cite{HHR82}, Lebzelter et al. \cite{LHH99}, Nowotny et al. \cite{Nowo05a}, \cite{Nowo05b}). The 
low excitation 2-0 transitions of $^{12}$CO are strong lines that form over a large part of the stellar atmospheres and also in the outflow of AGB stars. This makes the difference between line intensity variations in models P and PM 
understandable because the latter model has a larger extension and introduces an additional redshifted component to the line profile.

\begin{figure}
\resizebox{\hsize}{!}{\includegraphics[clip]{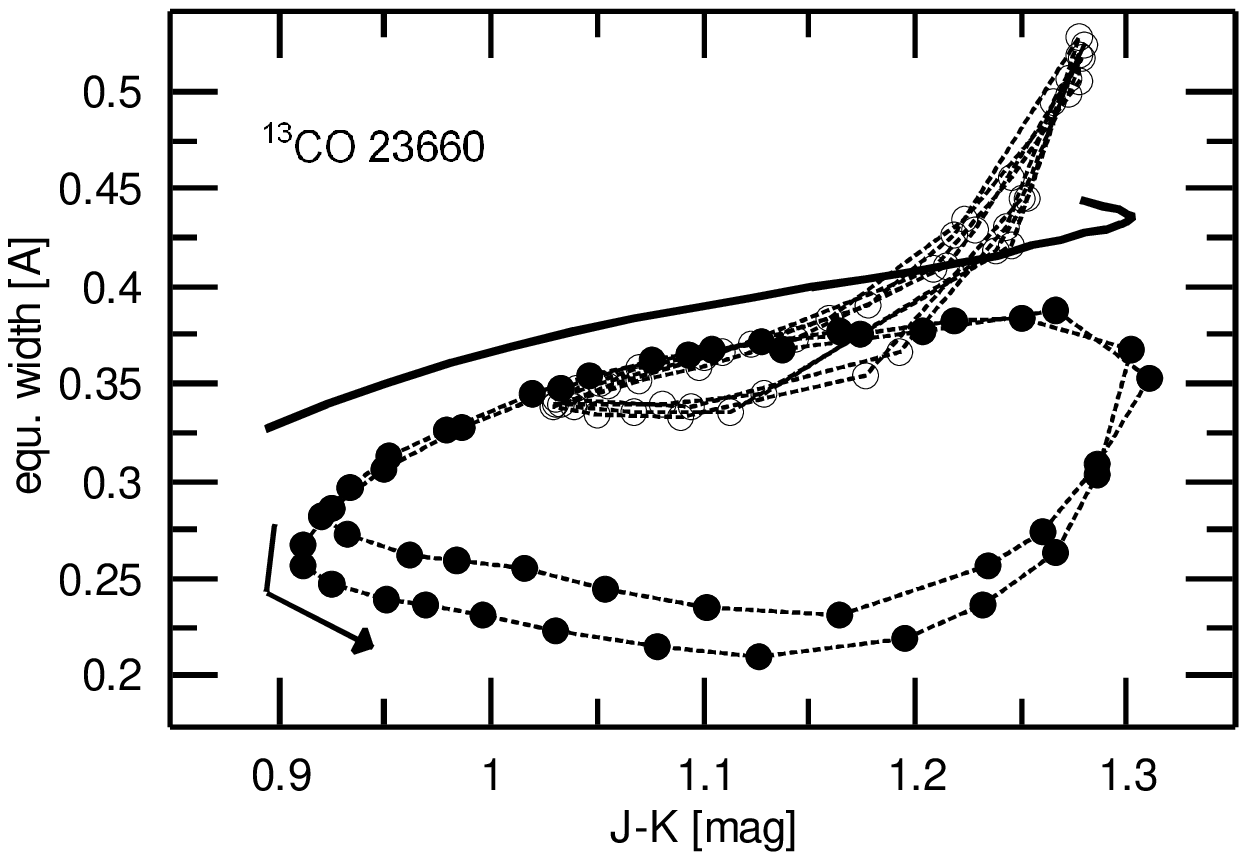}}
\caption{Same as Fig.\,\ref{f:Fe15536dyn} for the $^{13}$CO line at 23\,660\,{\AA}. Compare with Fig.\,\ref{f:13CO}.}
\label{f:13COdyn}
\end{figure}

For the $^{13}$CO line studied, we also find a behaviour deviating from the hydrostatic case (Fig.\,\ref{f:13COdyn}). 
Model P shows a trend very similar to the hydrostatic models over a large fraction of the pulsation cycle. The line intensity in model~PM follows the hydrostatic relation for about half the pulsation cycle, but for the other half the 
lines are much weaker. This cannot be accounted for by a change in log\,$g$ alone and would require an enormous change in the isotopic ratio. The occurrence of an emission component in the line profile is responsible for this 
weakening as in the case of the $^{12}$CO line. The emission is visible at several phases as a separate component, 
but is probably hidden by a general weakening of the line during a much larger part of the pulsation cycle. If we take out the emission component by artificially setting all values above the pseudo-continuum to the continuum level, the loop becomes less extended in the lower right corner of Fig.\,\ref{f:13COdyn}. 
Contrary to the case of model~P, a flattening of the loop cannot be achieved in this way.

In Fig.\,\ref{f:13COdyn}, we also see noticeable cycle-to-cycle differences, especially in the case of model
PM. This is an interesting result keeping in mind that our model atmosphere is modulated by a strictly sinusoidal 
variation in the inner boundary. This behaviour was previously noted for other output quantities 
(e.g. degree of dust condensation) from this kind of models (Nowotny et al.\,\cite{Nowo05a}, \cite{Nowo05b}).
For models based on self-excitation pulsation or real stars (see next section), even larger 
cycle-to-cycle differences can be expected (Lebzelter et al.\,\cite{LHA01}).

\section{Results from observations}\label{s:obsvsmod}
 
It is naturally it is a critical question whether the behaviour found for various lines in the dynamic models are also visible 
in observations. For the hydrostatic case, a number of successful comparisons between observed and synthetic 
photometry and spectra exist (e.g. Loidl et al. \cite{Loidl01}, Aringer et al. \cite{Aringer02}, Lebzelter et al. 
\cite{lz08}). For the dynamical case, it could also be shown that the models reproduce observations of C-rich stars 
reasonably well (e.g. Gautschy-Loidl et al. \cite{GaHJH04}, Nowotny et al. \cite{Nowo05b}).

Long time series of high resolution spectra of LPVs are rare. Here we use a time series of FTS spectra in the $H$- 
and $K$-band of the mira R\,Cas. This object is a typical mira variable in the solar neighbourhood with a period of 
431\,$^{\rm d}$ and a visual brightness amplitude up to 9\,magnitudes. At minimum light, the star reaches a spectral 
type of M10 (Lockwood \cite{Lockwood70}, Wyckoff \& Wehinger \cite{WW71}). Its dusty envelope has been used to derive 
a mass-loss rate of a few 10$^{-7}$\,$M_{\odot}$yr$^{-1}$ (Schutte \& Tielens \cite{ST89}; Weigelt \& Yudin 
\cite{WY01}). This agrees with an older mass-loss rate derived from thermal circumstellar CO emission (Loup 
et al. \cite{Loup93}). The atmospheric kinematics of R\,Cas was described by Hinkle et al.\,(\cite{HSH84}). 
These authors note that around minimum light the near infrared CO line profiles investigated by them seem to be 
strongly affected by blending of the CO lines with H$_{2}$O lines. Strong water absorption bands were also  
reported by Truong-Bach et al.\,(\cite{Truong99}) and Aringer et al.\,(\cite{Aringer02}) in the mid-infrared range. 
Clearly R\,Cas is described by parameters quite different from our models, and we wish to explicitly state that the 
dynamic model atmospheres (Table\,\ref{t:dmaparameters}) were not designed to resemble this specific object. 
We did not attempt to reproduce the spectrum of R~Cas. The main goal of this study is to identify the basic problems and features of abundance analysis in dynamical atmospheres. Our 
comparison shall be made on a qualitative basis only. The main reason for choosing this set of spectra is 
that to our knowledge it is the most extensive time series of high-resolution near-infrared spectra of an M-type mira 
available.

Our time series was obtained by one of us (KH) between August 1983 and September 1986 with the FTS spectrometer at 
the Kitt Peak National Observatory 4\,m Mayall telescope (Hall et al. \cite{Hall79}). It covers two complete light 
cycles of this star. A standard data analysis described extensively in previous papers (e.g. Hinkle et al.\,
\cite{HHR82}, \cite{HSH84}) was applied to the spectra. Additional analysis was performed 
within IRAF using the same 
procedures as in the case of the synthetic spectra.

\begin{figure}
\resizebox{\hsize}{!}{\includegraphics[clip]{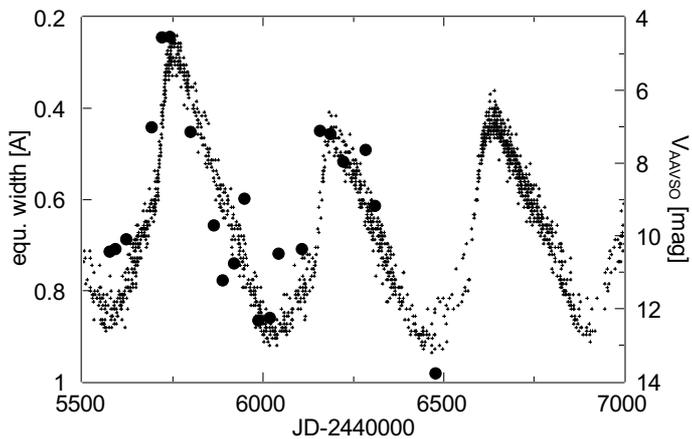}}
\caption{
Equivalent widths in [{\AA}] of the OH blend at 15570\,{\AA} (large dots, left y-axis) and the AAVSO light curve (small dots, right y-axis) for R\,Cas against time expressed in JD.}
\label{f:rcasphase}
\end{figure}

Phases within the pulsation cycle were attributed to the individual observations by using the AAVSO lightcurve (Fig.
\,\ref{f:rcasphase}). The figure clearly indicates that by chance our time series covers an untypical behaviour 
of the star with an outstandingly bright maximum during the first cycle and an unusually weak maximum in the second. 
The third maximum plotted represents the typical amplitude observed for this star. That this behaviour is reflected in an untypical atmospheric dynamics was demonstrated by Lebzelter et al. (\cite{LHA01}). This has to be kept in mind when discussing the line behaviour in R~Cas.

($J$--$K$) values obtained parallel to our time series do not exist. The atypical light change during that 
time does not allow us to use near-infrared light curves obtained during some other pulsation cycle (e.g. Kerschbaum et al. \cite{KLL01}, Nadzhip et al. \cite{nadzhip}). The only quantity directly available is the visual brightness from the AAVSO light curve. 
Assuming that ($J$--$K$) changes approximately in phase with the visual brightness, we use this indicator in our investigation. The $V$ value was derived from a polynomial fit to all AAVSO data. 
No correction for the difference between visual measurements and photometric $V$-band magnitudes was
applied, but the effect on the shape of the light curve should be very small.

\begin{figure}
\resizebox{\hsize}{!}{\includegraphics[clip]{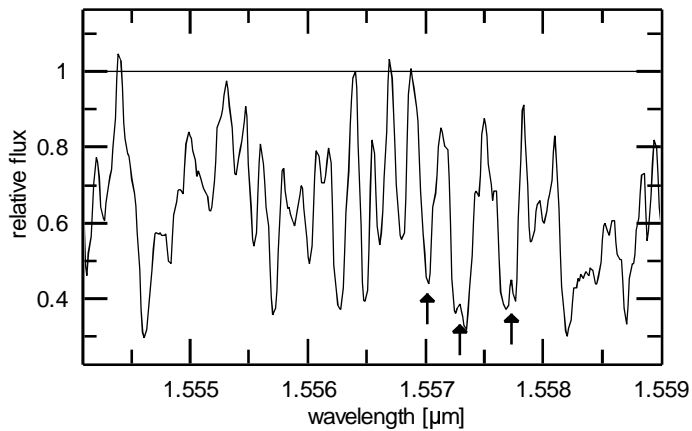}}
\caption{FTS spectrum of R~Cas around the OH lines obtained at JD~2\,445\,575 (corresponding to $\phi_{\rm V}$\,=\,0.59). The line indicates the chosen location of the pseudo-continuum. Arrows mark the three OH lines at 
15\,570, 15\,573, and 15\,576\,{\AA}, respectively (cf. upper right panel of Fig.\,\ref{f:linesstudied}). Note the clear indication of line doubling in the two stronger OH lines. The CO 3-0 band head can be seen on the right side of the plot.}
\label{f:FTSexample}
\end{figure}

We attempted to measure spectral features in the R~Cas data in a similar way 
(i.e. using the same width to determine the 
equivalent widths of lines) as in the synthetic spectra. We naturally had 
some difficulties in this case: the 
definition of the continuum is far more complicated in the observations than in the synthetic spectra, where emission 
peaks can be easily identified from the nominal continuum level. We illustrate our definition of the continuum in 
Fig.\,\ref{f:FTSexample}. As for the synthetic spectra (Sect.\,\ref{s:measuringwidths}), the observed spectra were shifted 
in wavelength (while preserving the Doppler-modified profiles) to allow an automatic measurement 
of equivalent widths. 
For comparison, we chose features that are only slightly contaminated 
by telluric absorption and assume that the influence 
of the Earth's atmosphere is negligible. Finally, our data are slightly affected by the instrument profile for which 
we also did not apply any corrections.

\begin{figure}
\resizebox{\hsize}{!}{\includegraphics[clip]{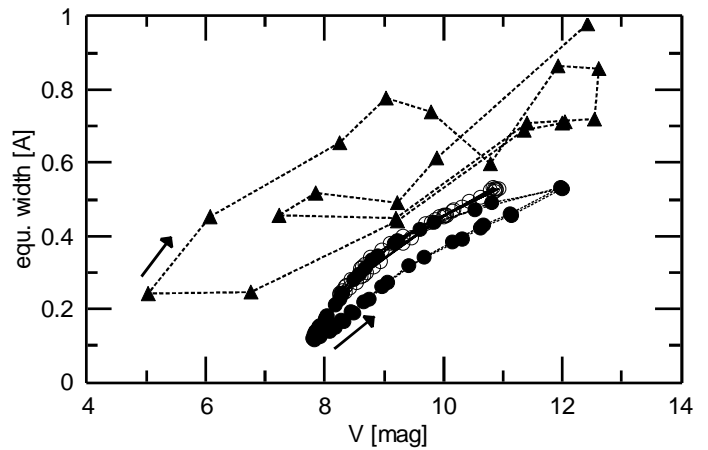}}
\caption{
Equivalent widths in [{\AA}] of the OH blend at 15570\,{\AA} measured from our time series of FTS spectra of the mira R\,Cas (triangles) against the corresponding $V$ values taken from the AAVSO light curve. 
The small arrow indicates the path from the first visual light maximum to minimum. For comparison, we also added 
the results from model P (open circles) and model PM (filled circles). Note that the model parameters are very 
different from the parameters of R Cas.}
\label{f:rcasOH}
\end{figure}

The most remarkable features in the case of the synthetic spectra are certainly the loops derived from the dynamic models, especially those models that produce an outflow of material. For a given near-infrared colour, two line strengths are found depending on the star being on the rising or falling part of the light change. R~Cas is obviously also a mira with a substantial mass loss. 
In Fig.\,\ref{f:rcasOH}, we show the behaviour of the OH line at 15\,570\,{\AA} versus visual brightness. 
Consecutive data points in the light cycle are connected by lines. Altering the continuum level by 2\,\% changes the  measured equivalent width by about 0.03\,{\AA}. Other spectral features checked by ourselves
show a similar behaviour. 

For an easier comparison, we also show the behaviour of the equivalent width with visual 
brightness derived from our models P and PM. The absolute visual magnitudes predicted by 
the models were shifted to
roughly coincide with the $V$-brightness of R~Cas. It is clear that the observed behaviour in R~Cas is in qualitative 
agreement with that found in our study using synthetic spectra. 
However, we note that the loop over the light cycle goes in the inverse direction (see the arrow in Fig.\,
\ref{f:rcasOH}). At the moment, the reason for this difference is not clear. Figure \ref{f:rcasphase} shows 
a strong correlation between the feature intensity and the visual brightness. 
The difference between the first and the 
second observed light cycle mentioned above is therefore 
clearly reflected in a corresponding change in the equivalent width. 
As a consequence, the loops resulting from these two light cycles (Fig.\,\ref{f:rcasOH}) have different 
extensions.

\begin{figure}
\resizebox{\hsize}{!}{\includegraphics[clip]{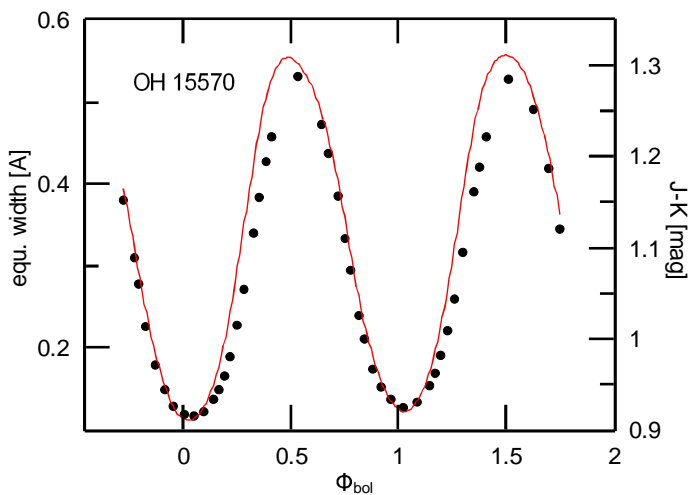}}
\caption{Equivalent widths in [{\AA}] of the OH blend at 15570\,{\AA} (solid symbols, left y-axis) and
($J$--$K$) colour (solid line, right y-axis) for model~PM against the bolometric phase. Phase 0 has been
assigned to the light maximum}
\label{f:OHphase}
\end{figure}

The occurrence of loops in a digram of equivalent width versus either ($J$--$K$) or $V$ corresponds to a different relation
between the two quantities on the rising and the falling branch of the light cycle. This can be seen in both 
Fig.\,\ref{f:OHphase} for the model case and Fig.\,\ref{f:rcasphase} for the observations.

\section{Conclusions and outlook}

The aim of our study has been to explore the possibilities and caveats of deriving stellar abundances from observed spectra of AGB stars that are strongly affected by pulsation and mass loss. We can summarise our findings as follows:

\begin{itemize}

\item  There is not a single phase, neither for model~P nor for model~PM, where all investigated features would be in agreement with a single hydrostatic model at the same time. This suggests that there is a fundamental problem in fitting the complete spectrum of a dynamical star with one hydrostatic model.

\item For several features, the dynamic models, especially model~PM, touch areas in the ($J$--$K$) versus 
equivalent width plane that seem to be unreachable by hydrostatic models even when modifying the chemical composition. Further tests are necessary to test the effects of e.g. metallicity on this finding.

\item Spectral peculiarities due to stellar dynamics seem to be less significant 
in the $H$-band than in the $K$-band. 

\item For dynamic atmospheres with mass loss (represented by model~PM), the rising, post-minimum branch seems to be 
more promising for deriving element abundances with the help of hydrostatic models. This is, however, untrue for some of the features studied (namely the iron lines, OH 15540 and $^{12}$CO 23137).

\item For dynamical atmospheres without mass loss (model~P), the loops are less extended and the detected trends of equivalent width with colour seem to resemble the hydrostatic case in many respects. For several features, 
we find an offset between the relation for model~P and the hydrostatic reference case with log\,$g=-$0.25. This offset, however, is not the same for all features.  

\item Including the emission lines in the measurement of the strength of a spectral feature leads to an extension of 
the loop completed in the colour versus equivalent width plane. For high resolution spectroscopy, it is therefore 
better to exclude any clearly separated emission component when deriving the stellar abundance. If low resolution spectra were used, in which the emission component cannot be detected separately, we expect to see
extensive loops such as those found in this study.

\item Our results are qualitatively  similar to the findings of Scholz (\cite{Schol92}) and others. 

\end{itemize}

We performed some tests to see if a combination of various spectral features (e.g. a ratio of the equivalent widths of two lines) could reduce the complexity of the variation with colour. But as the size and slope of the loops are 
different for each of the features observed, none of the combinations were found to be useful.

Deriving stellar abundances for pulsating stars without an outflow (simulated in this study by model~P with a bolometric light amplitude of 0.86$^{\rm mag}$) may well be possible if we take into account a set of offsets that still need to be determined. Abundance determinations for these kind of stars performed 
so far are probably affected by systematic error, depending on the feature used.

If we wish to derive stellar abundances from dynamic atmospheres, we have to search for features that are far 
more sensitive to changes in the chemical composition than to dynamical effects. An example would be the $^{12}$CO 
3-0 band head. As described above, the loop (Fig.\,\ref{f:COband_dyn}) covers a range in the colour vs. equivalent 
width diagram similar to a change in log\,$g$ by 0.5. This is clearly less than the change from C/O$=$0.48 to 
C/O$=$0.25 (Fig.\,\ref{f:CO_band}). Therefore we have a good chance of determining whether, e.g., the surface 
abundances of an AGB star have already been modified from the first dredge-up composition by the third dredge-up 
(e.g. Lattanzio \& Wood \cite{LW04}). Having measurements of different phases would naturally help us 
to improve the accuracy of the result.

We emphasize that our approach includes some simplifications that need to be explored in more detail. We did not test 
the effect of a variation in metallicity. Furthermore, a change in C/O may have a more complex effect 
on dynamic atmospheres than known from the hydrostatic case. Finally, a comparison with observed time series 
of a mira reveals that some features are in common with the model (e.g. the occurrence of loops), while other aspects are different (e.g. the dependency on phase). At present, it is not clear if this is due to a general problem in the 
models or a dependence of these characteristics on the parameters of the model (note that R\,Cas is very  
different from the models we used in our analysis). However, we point out that the values found for the equivalent 
widths in R\,Cas are in good agreement with what we would expect from an extension of the detected trends towards redder colours.

In a forthcoming paper, we will continue our comparison of high resolution synthetic spectra from dynamical and 
hydrostatic model atmospheres to identify the most well-suited features for measuring stellar parameters and abundances for cool large amplitude variables. We also plan to extend the study to dynamic models with different 
composition. Finally, we intend to directly compare the observed spectra of AGB variables in the globular
cluster 47 Tuc with the results of hydrostatic and dynamic model atmospheres.

\begin{acknowledgements}
This work has been supported by the Austrian Science Fund (P18171-02, P18939-N16, P20046-N16 and P21988-N16). 
BA acknowledges funding by the contract ASI-INAF I/016/07/0.
We acknowledge with thanks the variable star observations from the AAVSO International Database contributed by observers worldwide and used in this research.
\end{acknowledgements}



\begin{thebibliography}{}

\bibitem [1989] {AndeG89} 
Anders, E., \& Grevesse, N. 1989, Geochimica et Cosmochimica Acta 53, 197

\bibitem [1999] {AHWHJ99}
Aringer, B., H\"ofner, S., Wiedemann, G., et al. 1999, A\&A, 342, 799

\bibitem [2000] {Aring00}
Aringer, B. 2000, PhD thesis, University of Vienna, Austria

\bibitem[2002]{Aringer02} 
Aringer, B., Kerschbaum, F., \& J{\o}rgensen, U.G. 2002, A\&A 395, 915

\bibitem [2008] {AriNH08}
Aringer, B., Nowotny, W., \& H\"ofner, S. 2008, in Perspectives in Radiative Transfer and Interferometry, ed. S., Wolf, F., Allard, \& Ph., Stee, EAS Publ. Ser., 28, 67

\bibitem [2009]{Aringer09}
Aringer, B., Girardi, L., Nowotny, W., Marigo, P., \& Lederer, M.T. 2009, A\&A, 503, 913

\bibitem [1991] {BasSW91}
Baschek, B., Scholz, M., Wehrse, R. 1991, A\&A, 246, 374

\bibitem [1989] {BessS89}
Bessell, M.S., \& Scholz, M. 1989, 
in \textit{Evolution of Peculiar Red Giants}, IAU Coll. 106, ed. H.R. Johnson, B. Zuckerman, Cambridge University Press, p.67

\bibitem [1996] {BesSW96}
Bessell, M.S., Scholz, M., \& Wood, P.R. 1996, A\&A, 307, 481

\bibitem [1988] {Bowen88}
Bowen, G.H. 1988, ApJ, 329, 299

\bibitem [1997] {CarrG97}
Carretta, E., \& Gratton, R.G. 1997, A\&AS, 121,95

\bibitem [1990] {CH90}
Castelli, F., \& Hack, M. 1990, Mem. Soc. Astron. Ital., 61, 595

\bibitem [2007] {CrSLA07}
Cristallo, S., Straniero, O., Lederer, M.T., \& Aringer, B. 2007, ApJ, 667, 489

\bibitem [1992] {FleGS92}
Fleischer, A.J., Gauger, A., Sedlmayr, E. 1992, A\&A, 266, 321

\bibitem [2004] {GaHJH04}
Gautschy-Loidl, R., H\"ofner, S., J{\o}rgensen, U.G., \& Hron, J. 2004, A\&A,
422, 289 

\bibitem [1994] {GoorC94}
Goorvitch, D., \& Chackerian, C.Jr. 1994, ApJS, 91, 483

\bibitem [1994] {GrevS94} 
Grevesse, N., \& Sauval, A.J. 1994, in \textit{Molecules in the Stellar Environment}, IAU Coll. 146, ed. U.G. J{\o}rgensen, Lecture Notes in Physics 428, Springer, p.~196

\bibitem[1979]{Hall79} 
Hall, D.N.B., Ridgway, S.T., Bell, E.A., \& Yarborough, J.M. 1979, Proc. SPIE 172, 121       

\bibitem[1982]{HHR82}
Hinkle, K.H., Hall, D.N.B., \& Ridgway, S.T. 1982, ApJ, 252, 697

\bibitem[1984]{HSH84}
Hinkle, K.H., Scharlach, W.W.G., \& Hall, D.N.B. 1984, ApJS, 56, 1

\bibitem [1998] {HofSW98}
Hofmann, K.-H., Scholz, M., \& Wood, P.R. 1998, A\&A, 339, 846

\bibitem [2003] {HoGAJ03}
H\"ofner, S., Gautschy-Loidl, R., Aringer, B., \& J{\o}rgensen, U.G. 2003, 
A\&A, 399, 589 

\bibitem [2007] {Hoefn07}
H\"ofner, S. 2007, in \textit{Why Galaxies Care About AGB Stars}, ed. F. Kerschbaum, C. Charbonnel, B. Wing., ASP Conf. Ser., 378, 145

\bibitem [2008] {Hoefn08}
H\"ofner, S. 2008, A\&A, 491, L1

\bibitem [1997] {Jorge97}
J{\o}rgensen, U.G. 1997, in \textit{Molecules in Astrophysics: Probes and Processes}, ed. E.F. van Dishoek, IAU Symp., 178, p.441

\bibitem [1993] {JorgJ93}
J{\o}rgensen, U.G., \& Jensen, P. 1993, J. Mol. Spectrosc., 161, 219

\bibitem [2001] {JoJSA01}
J{\o}rgensen, U.G., Jensen, P., S{\o}rensen, G.O., \& Aringer, B. 2001, A\&A, 372, 249

\bibitem[2001]{KLL01} 
Kerschbaum, F., Lebzelter, T., \& Lazaro, C. 2001, A\&A, 375, 527

\bibitem[2000]{Kupka00}
Kupka, F.G., Ryabchikova, T.A., Piskunov, N.E., et al. 2000, BaltA, 9, 590

\bibitem[2004] {LW04}
Lattanzio, J.C., \& Wood, P.R. 2004, in: Asymptotic Giant Branch Stars, eds.
Habing, H.J., Olofsson, H., Springer, New York, p. 23

\bibitem[1999]{Lebzelter99}
Lebzelter, T. 1999, A\&A, 351, 664

\bibitem[1999]{LHH99}
Lebzelter, T., Hinkle, K.H., \& Hron, J. 1999, A\&A, 341, 224

\bibitem[2001]{LHA01}
Lebzelter, T., Hinkle, K.H., \& Aringer, B. 2001, A\&A, 377, 617

\bibitem [2005] {LebzW05}
Lebzelter, T., \& Wood, P.R. 2005, A\&A, 441, 1117

\bibitem [2005] {LWHJF05}
Lebzelter, T., Wood, P.R., Hinkle, K.H., Joyce, R.R., \& Fekel, F.C. 2005, A\&A, 432, 207

\bibitem [2006] {LPHWB06}
Lebzelter, Th., Posch, Th., Hinkle, K., et al. 2006, ApJ, 653, L145

\bibitem[2008]{lz08} 
Lebzelter, T., Lederer, M.T., Cristallo, S. et al. 2008, A\&A, 486, 511       

\bibitem [2009] {LedeA09}
Lederer, M.T., \& Aringer, B. 2009, A\&A, 494, 403

\bibitem[2001]{Loidl01} 
Loidl, R., Lancon, A., \& J{\o}rgensen, U.G. 2001, A\&A, 371, 1065

\bibitem[1970]{Lockwood70}
Lockwood, G.W. 1970, ApJ, 160, L47

\bibitem[1993]{Loup93}
Loup, C., Foreville, T., Omont, A., Paul, J.F. 1993, A\&AS, 99, 271

\bibitem[2007]{MWSLH07}
McSaveney, J.A., Wood, P.R., Scholz, M., et al. 2007, MNRAS, 578, 1089

\bibitem[2001]{nadzhip} 
Nadzhip, A.E., Tatarnikov, A.M., Shenavrin, V.I., et al. 2001, Astronomy Letters 27, 376       

\bibitem [2005] {Nowot05}
Nowotny, W. 2005, \textit{The Dynamic Atmospheres of Red Giant Stars. Spectral
Synthesis in High Resolution}, PhD thesis, University of Vienna, Austria 

\bibitem[2005a]{Nowo05a}
Nowotny, W., Aringer, B., H\"ofner, S., Gautschy-Loidl, R., Windsteig, W. 2005a,
A\&A, 437, 273

\bibitem [2005b] {Nowo05b}
Nowotny, W., Lebzelter, T., Hron, J., \& H\"ofner, S. 2005b, A\&A, 437, 285

\bibitem [2010] {NowAH09}
Nowotny, W., Aringer, B., \& H\"ofner, S. 2010, A\&A, in press (AA/2009/11899)

\bibitem [2009] {PAHNS09}
Paladini, C., Aringer, B., Hron, J., et al. 2009, A\&A 501, 1073

\bibitem [2005] {RJBCB05}
Rothman, L.S., Jacquemart, D., Barbe, A., et al. 2005, \\
\jqsrt, 96, 139

\bibitem [1992] {Schol92}
Scholz, M., 1992, A\&A, 253, 203

\bibitem [2002] {Smith02}
Smith, V.V., Hinkle, K.H., Cunja, K., et al. 2002, ApJ, 124, 3241

\bibitem[1989]{ST89} 
Schutte, W.A., \& Tielens, A.G.G.M. 1989, ApJ, 343, 369

\bibitem[1999] {Truong99}
Truong-Bach, Sylvester, R.J., Barlow, M.J. et al. 1999, A\&A, 345, 925

\bibitem[2001] {WY01}
Weigelt, G., \& Yudin, B.F. 2001, Astronomy Reports, 45, 510

\bibitem[2000] {Winters00}
Winters, J.M., Keady, J.J., Gauger, A., \& Sada, P.V. 2000, A\&A, 359, 651

\bibitem[1997] {WiFLS97}
Winters, J.M., Fleischer, A.J., Le Bertre, T., \& Sedlmayr, E. 1997, A\&A, 326, 305

\bibitem [2006] {Woitk06b}
Woitke, P. 2006, 
A\&A, 460, L9

\bibitem[1971]{WW71}
Wyckoff, S., Wehinger, P. 1971, ApJ, 164, 383

\bibitem [1988] {Yorke88}
Yorke, H.W. 1988, in \textit{Radiation in moving gaseous media}, ed. Y. Chmielewski, T. Lanz, 18$^{th}$ Advanced Course of the Swiss Society of Astrophysics and Astronomy (Saas-Fee Course), p.193

\end{thebibliography}
\end{document}